\documentclass[12pt,a4paper]{article}
\usepackage[utf8]{inputenc}
\usepackage[russian]{babel}
\usepackage[OT1]{fontenc}

\usepackage{cite}

\usepackage{amsmath}
\usepackage{amsfonts}
\usepackage{amssymb}

\usepackage{feynmp}
\usepackage{hyperref}

\usepackage{appendix}

\newcommand{\uunderline}[1]{\underline{\underline{#1}}}
\newcommand{\pd}{\partial}
\newcommand{\ket}[1]{\left\lvert #1 \right\rangle}

\usepackage{authblk}

\title{Базовые проблемы консервативных подходов к квантовой теории гравитации}
\author[1,2,3]{Б.Н. Латош}
\affil[1]{Университет Сассекса, Brighton, BN1 9QH, United Kingdom}
\affil[2]{Государственный Унивреситет ``Дубна'', Университетская 19, Дубна, 141982, Россия}
\affil[3]{Лаборатория теоретической физики им. Н.Н.Боголюбова, ОИЯИ, Дубна, 141980б Россия}
\affil[ ]{\textit{latosh@theor.jinr.ru}}
\date{}

\begin{document}

\maketitle

\begin{abstract}
  Представлен обзор наиболее базовых проблем, возникающих при применении стандартных программ создания квантовой теории к гравитации. Большая часть работы посвящена проблемам пертурбативного квантования, основанного на формализме континуального интеграла. Обсуждены проблемы программы канонического квантования в контексте её применения к теории гравитации. Представлено краткое обсуждение применения методов эффективной теории поля в квантовой гравитации.
  
  {~}

\end{abstract}

\section{Введение}\label{introduction}

Создание квантовой теории гравитации по праву считается одной из наиболее фундаментальных задач современной теоретической физики. Она восходит к началу прошлого века и возникает в одно время с общей теорией относительности (ОТО). Сам Эйнштейн понимал необходимость создания отдельной теории гравитационного взаимодействия для квантовых систем, так, уже в 1916 году \cite{Einstein:1916cc} он писал следующее\footnote{``Тем не менее, атомы, следуя законам движения электронов, должны излучать не только электромагнитную, но также и гравитационную энергию, пусть и в малых количествах. Так как в действительности в природе этого может не происходить, похоже, квантовая теория должна будет изменить не только уравнения Максвелла, но и уравнения гравитационной теории.''}:
\begin{quote}
  ``Gleichwohl m\"u{\ss}ten die Atome zufolge der inneratomischen Elek\-tro\-nen\-be\-we\-gung nicht nur elektromagnetische, sondern such Gra\-vi\-ta\-ti\-ons\-ener\-gie ausstrahlen, wenn auch in winzigem Betrage. Da dies in Wahrheit in der Natur nicht zutreffen d\"urfte, so seheint es, da\ss{} die Quantentheorie nicht nur die Maxwellsehe Elektrodynamik, sondern auch die neue Gravitationstheorie wird modifziren m\"ussen.''
\end{quote}

Развитие представлений о квантовой теории и её применение для описания гравитационного взаимодействия привело к созданию множества сложнейших техник и моделей таких как AdS/CFT соответствие, модели с дополнительным измерениями, петлевая квантовая гравитация, супергравитация и теория суперструн. Подробный обзор исторического развития идей о квантовой гравитации вместе с обсуждением генеалогии современных подходов приводится в \cite{RovelliCarlo2004Qg,Rovelli:2000aw}.

Колоссальный прогресс, достигнутый в этой области, как ни парадоксально, довольно плохо описан в современной литературе. Значительная часть книг, так или иначе посвящённых проблеме квантовой гравитации, а именно \cite{RovelliCarlo2004Qg,Smolin:2003rk,Ashtekar:1991hf,Thiemann:2007zz,ZwiebachBarton2004Afci,Becker:2007zj,Green:2012oqa,Green:2012pqa,Polchinski:1998rq,Polchinski:1998rr,smolin2014three}, не освещает наиболее базовые проблемы, препятствующие построению квантовой теории гравитации стандартными методами квантовой теории поля (КТП). На сколько нам известно, в контексте современной квантовой поля эти проблемы подробно освещены лишь в книгах \cite{RN18,RN19,RovelliCarlo2004Qg,Vergeles_Gravity} из которых \cite{RN18,RN19,RovelliCarlo2004Qg} на русском языке не издавались.

Цель этого обзора -- осветить базовые проблемы, препятствующие построению квантовой теории гравитации по полной аналогии с известными квантовыми теориями поля, такими как квантовая электродинамика или квантовая хромодинамика. А именно, мы обсудим проблемы, возникающие при применении формализма КТП к теории гравитации в четырёхмерном пространстве-времени. Отметим также, что все теории, являющиеся кандидатами на роль квантовой теории гравитации, имеют свои фундаментальные проблемы, часть из которых принципиально отличается от проблем, обсуждаемых в данном обзоре.

Мы проследуем по двум наиболее простым и консервативным программам квантования и покажем, что они не могут быть признаны удовлетворительными. В разделе \ref{the_overview} мы представим две простейшие программы квантования гравитационного взаимодействия -- пертурбативную и каноническую. В разделе \ref{quantum_GR} мы обсудим применение пертурбативного квантования к ОТО. Пертурбативный метод квантования основывается на использовании функционального интеграла и метода квантования во внешнем поле, что позволяет построить квантовую модель малых метрических возмущений на фоне плоского пространства-времени. В разделе \ref{quantum_GR_beyond_tree_level} мы обсудим проблемы пертурбативного квантования теории гравитации, а именно, неперенормируемость квантовой ОТО, духовые состояния в спектре квантовых моделей гравитации и проблему космологической постоянной. В разделе \ref{nonperturbative_quantum_gravity} мы обсудим канонический подход к квантованию гравитации и возникающие при этом проблемы. Наконец, в разделе \ref{effective_quantum_gravity} мы обсудим эффективную квантовую гравитацию -- подход, основанный на методе эффективной теории поля, позволяющую согласованно описать влияние квантовых эффектов на макроскопическое проявление гравитационного взаимодействия. В разделе \ref{conclusion} мы суммируем все проблемы, возникающие при попытке консервативного построения квантовой теории гравитационного взаимодействия, и обсудим их роль в современных исследованиях квантовой гравитации.

В разделах \ref{quantum_GR}, \ref{quantum_GR_beyond_tree_level} и \ref{effective_quantum_gravity} мы используем метрику с сигнатурой $(+---)$; в разделе \ref{nonperturbative_quantum_gravity} мы используем сигнатуру $(-+++)$, чтобы трёхмерные расстояния имели положительную длину. Во всех разделах мы используем тензоры Римана и Риччи определённые следующим образом:
\begin{align}
  \begin{split}
    R_{\mu\nu}{}^\alpha{}_\beta &= \pd_\mu \Gamma^\alpha_{\nu\beta} -\pd_\nu \Gamma^\alpha_{\mu\beta} + \Gamma^\alpha_{\mu\lambda}\Gamma^\lambda_{\nu\beta} - \Gamma^\alpha_{\nu\lambda}\Gamma^\lambda_{\mu\beta}~, \\
    R_{\mu\nu} &= R_{\sigma\mu}{}^\sigma{}_\nu~.
  \end{split}
\end{align}
Греческие индексы принимают значения от $0$ до $3$. Латинские индексы используются только в разделе \ref{nonperturbative_quantum_gravity}, принимают значения от $1$ до $3$ и связываются с координатами $x^a$ на пространственно-подобных слоях.

\section{Консервативные подходы к квантовой гравитации}\label{the_overview}

В контексте квантовой теории гравитации прежде всего необходимо обсудить вопрос о том, какой именно смысл вкладывается в процедуру квантования. Этот вопрос важен с методологической точки зрения, так как он определяет способ создания квантовой теории. Квантовая и классическая теории предоставляют различные способы описания физической реальности. В рамках классической физики невозможно описать явления, связанные с нелокальными свойствами материи, такими как дифракцию электронов \cite{thomson_diffraction_1927,Davisson317}, неравенства Белла \cite{BRAUNSTEIN199022} и существование ``квантового ластика'' (quantum eraser)\cite{Walborn:2002zza}. Поэтому квантовая теория считается более фундаментальной, в кто время как классическая играет роль приближённого описания. Однако, классические модели полей более интуитивно понятны и часто используются для создания квантовых моделей.

В результате такого подхода может сложиться ложное впечатление, что для построение квантовой модели необходимо использовать какую-либо классическую модель. Это убеждение ложно, что может быть показано на примере квантовой хромодинамики. Несмотря на то, что возможно создать модель нескольких классических полей, имеющих калибровочную $SU(3)$ симметрию, невозможно полностью восстановить феноменологию квантовой хромодинамики. Препятствует этому конфайнмент кварков -- квантовое явление, не имеющее аналогов в классической физке \cite{PhysRevD.10.2445,Greensite:2011zz}. Из-за конфайнмента кварки не могут оказаться в низкоэнергетическом свободном состоянии и могут лишь формировать связанные состояния, например, адроны. По этой причине в режиме низких энергий кварки не смогут сформировать классическое кварковое поле, а будут формировать связные состояния. В этом смысле квантовая хромодинамика не имеет классического предела, так как классическое (низкоэнергетическое) поле кварков не может реализоваться в природе.

Аналогичная ситуация может реализовываться и в гравитационном секторе. Подавляющее большинство эмпирических данных хорошо согласуется с ОТО, однако, это не является достаточным основанием полагать, что квантовая теория гравитации представляет из себя результат применения некой формальной процедуры квантования к ОТО. Наиболее известным примером подобного подходя являются струнные модели. В их рамках фундаментальное описание всех физических взаимодействий даётся в терминах суперструн -- объектов никак не связанных ни с классической физикой, ни с ОТО. Сама же ОТО, как считается, возникает в естественным образом в пределе низких энергий. Таким образом, нет основания полагать, что корректная квантовая теория гравитации с необходимостью должна основываться на ОТО.

Следуя этой логике необходимо выделять два типа квантовых моделей гравитации. Первые, которым посвящён данный обзор, следует называть консервативными, так как они исходят из предположения о том, что квантовая теория гравитации может быть получена из ОТО (или другой классической модели гравитации) при помощи формальной процедуры квантования. Модели другого типа, которые лишь упоминаются в этом обзоре, исходят из того, что квантования теория гравитации должна быть сформулирована на основании принципиально новых физических объектов. К моделям такого типа относятся, например, упомянутые выше струнные модели или петлевая квантовая гравитация. Именно базовые проблемы консервативных подходов привели к возникновению неконсервативных моделей и, в конечном счёте, к формированию современных воззрений на проблемы квантовой теории гравитации.

Существует два простейших консервативных способа создания квантовых полевых моделей. Первый мы будем называть пертурбативным квантованием, и в его основу положим технику функционального интеграла. Второй подход мы будем называть каноническим, и в его рамках будем рассматривать алгебраическое описание квантовых систем, данное через алгебру операторов и свойства пространства состояний. Эти подходы основываются на минимальном числе предположений и предоставляют простые алгоритмы для создания квантовых моделей по классическим аналогам. Кроме того, они отражают две комплементарные тенденции в развитии представлений о квантовой гравитации.

Суть пертурбативного подхода заключается в создании квантовой теории малых возмущений фоновых физических полей. Этот подход получил наибольшее распространение в современной физике из-за простоты применения к прикладным задачам. Пертурбативный подход формулируется при помощи техники функционального интеграла и использует метод квантования во внешнем поле \cite{DeWitt:1967yk,DeWitt:1967ub,DeWitt:1967uc,zinn-justin_path_2011,zinn-justin_quantum_2012,KlubergStern:1974xv,KlubergStern:1975hc,Abbott:1983zw,Hart:1984jy,Boulware:1980av,Denner:1994xt}.

Квантование осуществляется согласно следующему алгоритму. Прежде всего, задаётся набор физических полей $\{\Psi\}$ и фундаментальное (микроскопическое) действие $\mathcal{A}[\Psi]$, определяющее динамику всей системы. Каждое физическое поле разбивается на две части: одна описывает фоновую структуру поля $\overline\Psi$, другая -- малым возмущениям этого фона $\psi$. Наконец, задаётся генерирующий функционал квантовой системы (приводимый здесь с точностью до бесконечного нормирующего множителя):
\begin{align}
  Z[~\overline{\Psi}~] &\overset{\text{def}}{=} \int\mathcal{D}[\psi] \exp\left[ i \, \mathcal{A}\left[~\overline{\Psi}+\psi\right]\right] \label{action_s_1} \\
  &= \int\mathcal{D}[\psi] \exp\left[ i\, \mathcal{A}[\overline{\Psi}] + i\,\cfrac{\delta \mathcal{A}[~\overline\Psi~]}{\delta\overline{\Psi}}\, \psi + i\, \cfrac{\delta^2 \mathcal{A} [~\overline\Psi~]}{\delta\overline\Psi^2}\, \psi^2 + O(\psi^3)  \right] .\nonumber
\end{align}
Вся информация о поведении физической системы содержится в этом генерирующем функционале. Получить её можно при помощи вычисления многочастичных функций, следуя стандартному алгоритму \cite{zinn-justin_path_2011,zinn-justin_quantum_2012,RN14}.

Фоновая часть полей $\overline{\Psi}$ описывает макроскопическое состояние системы, в котором распространяются малые квантовые флуктуации $\psi$. Если фон $\overline\Psi$ в отсутствии внешних источников является решением классических уравнений поля, то есть доставляет минимум микроскопическому действию, то его следует интерпретировать как основное состояние полей данной модели. Его возмущения должны быть интерпретированы как фундаментальные возмущения основного состояния полей, то есть как фундаментальные частицы-переносчики поля. Однако, если фон $\overline\Psi$ есть решение классических уравнений в присутствии внешних источников, то его следует интерпретировать как конфигурацию поля равновесную со внешними источниками. В таком случае возмущения поля будут описывать возмущения этой равновесной конфигурации поля и должны быть интерпретированы, скорее, как квазичастицы, возникающие в среде, поляризованной внешними источниками. Такой подход позволяет изучать квантовые системы во внешних классических полях, например, электромагнитное взаимодействие электронов в поле ядра атома водорода \cite{Schwinger:1951nm}, или в поле кристаллической решётки твёрдого тела \cite{KittelCharles2005Itss}. Аналогично можно поставить задачу об изучении квантовой хромодинамики на нетривиальном фоне \cite{Novikov:1983gd,Brower:1981vt}. Подчеркнём отдельно, что при пертурбативном квантовании фоновое поле $\overline\Psi$ рассматривается как внешний фактор, не зависящий от динамики самой системы. Обсуждение обратной реакции классического фона, вызванной динамикой квантовой системы, требует особого изучения и выходит за рамки этого обзора.

Эту логику проще всего продемонстрировать на примере электродинамики. При построении квантовой электродинамики на фоне пустого про\-стран\-ства\--вре\-ме\-ни мы интерпретируем малые квантовые возмущения электромагнитного поля как фотоны, которые являются фундаментальными частицами. Из-за линейной природы электромагнитного поля мы можем применить точно такую же интерпретацию и к электродинамике во внешнем поле. Например, при изучении электродинамики в поле ядра атома водорода мы также интерпретируем возмущения электромагнитного поля как фотоны. В теориях же с нелинейным взаимодействием такая интерпретация неприменима. Например, в рамках квантовой теории электронов в твёрдом теле невозможно интерпретировать все возмущения поля электронов как фундаментальные частицы. А именно, можно ввести состояния возмущения поля электронов, называемые ``дырками'', которые не являются фундаментальными частицами. Аналогично можно ввести состояния возмущения решётки твёрдого тела, называемые ``фононами'', которые также нельзя считать фундаментальными частицами \cite{KittelCharles2005Itss}.

Канонический подход к квантовой теории основывается на изучении алгебры операторов, построенной по алгебре Пуассона классической системы \cite{Dirac:1925jy,DiracPaulAdrienMaurice1958TPoQ,RN16}. Квантовая теория описывает физические системы в терминах состояний $\ket{\text{state}}$ и операторов $\hat\Psi$, подчинённых алгебре, образованной коммутаторами $[\cdot,\cdot]$. В полной мере описать квантовую систему можно, если задать оператор Гамильтона $\hat{H}$, соответствующий генератору сдвига по времени, так что для любого оператора $\hat{F}$ будет выполняться уравнение Гейзенберга:
\begin{align}
  i\, \cfrac{\pd}{\pd t}\hat{F}=[\hat{H},\hat{F}]~.
\end{align}
Уравнение Гейзенберга позволяет ввести оператор эволюции системы $\hat{S}(t,t')$ (или $S$-матрицу), удовлетворяющий уравнению Гейзенберга по первой переменной:
\begin{align}
  i\, \cfrac{\pd}{\pd t} ~ \hat{S}(t-t') = [\hat{H}(t), \hat{S}(t-t')]~.
\end{align}
В таком подходе чтобы полностью описать физическую систему необходимо задать алгебру операторов и либо оператор Гамильтона, либо оператор эволюции. Этот метод может быть применён для решения практических задач, если определить оператор эволюции через плотность лагранжиана:
\begin{align}
  \hat{S}(t,t')\overset{\text{def}}{=}\exp\left[ ~i \int\limits_{t'}^t dt ~ \int d^3V ~\hat{\mathcal{L}} \right].
\end{align}
Тут под $dV$ обозначает инвариантный элемент объёма, а под $\hat{\mathcal{L}}$ -- плотность лагранжиана, определённую через квантовые операторы.

Подчеркнём, что в рамках этого подхода нет необходимости ни вводить фон, по отношению к которому будут определены малые возмущения, ни устанавливать какой бы то ни было малый параметр. Более того, операторы рождения и уничтожения состояний (вне зависимости от того, описывают эти состояния частицы или квазичастиц) не ограничены малыми возмущения. Необходимость ограничиться лишь малыми возмущениями связана с математическими свойствами рядом операторов. Оператор эволюции традиционно определяется через бесконечный ряд операторов. Для того, чтобы это формальное выражение сходилось, необходимо вводить ограничения на операторы взаимодействия, что физически эквивалентно изучению лишь малых возмущений и возвращает нас к пертурбативному описанию. Однако, это ограничение никак не связано ни с самим методом, ни с его физическим содержанием, а лишь с математическими требованиями, обеспечивающими формальную сходимость рядов.

Канонических подход к квантованию гравитации приводит к ряду специфических проблем. В отличии от пертурбативного подхода, осложнения возникают как только мы пытаемся применить метод канонического квантования. Теория гравитации является калибровочной теорией, в которой роль калибровочных преобразований играет смена координат \cite{Dirac:1958jc}. Поэтому аналогично другим калибровочным системам теория гравитации описывает динамическую системы со связями \cite{Arnowitt:1962hi,Dirac:1950pj,Prohorov_Shabanov_Hamiltonian_Systems}. Наличие калибровочной симметрии приводит к сложности описания пространства физически состояний теории, так как базис, построенный на собственных состояниях оператора поля, переопределен. Отдельную фундаментальную проблему представляет введение времени в каноническую теории гравитации, так как согласно нашим представлениям время не может быть параметром эволюции внешним по отношению к системе, а должно возникать динамически \cite{kuchar_time_2011}. Наконец, наиболее простой подход к каноническому квантованию гравитации имеет технические проблемы, связанные с тем, что уравнение Уилера-деВитта -- аналог уравнения Шрёдингера -- имеет пространство решений, на котором нельзя ввести положительную норму. Совокупность этих факторов не позволяет признать приемлемы современный канонический подход к квантованию гравитации.

Таким образом, два наиболее простых подхода к квантовому описанию гравитации содержат ряд недостатков. Как мы покажем далее, недостатки этих методов являются критически важными и не позволяют построить квантовую теорию гравитации имеющимися стандартными методами.

\section{Пертурбативное квантование гравитации}\label{quantum_GR}

Будем следовать программе пертурбативного квантования, приведённой в предыдущем разделе, и построим пертурбативную квантовую ОТО. Рассмотрим возмущения метрики $h_{\mu\nu}$ над фоновым пространством-временем $\bar{g}_{\mu\nu}$. Метрику, описывающую совокупность фона и возмущений, зададим следующей формулой:
\begin{align}\label{metric_with_perturbations}
  g_{\mu\nu} = \bar{g}_{\mu\nu} + \kappa h_{\mu\nu}~.
\end{align}
Размерная константа $\kappa$ определёна через постоянную Ньютона:
\begin{align}
  \kappa^2=32 \pi G ~.
\end{align}
Подчеркнём, что выражение \eqref{metric_with_perturbations} не является разложением в ряд и даёт точную связь между полной метрикой $g_{\mu\nu}$ и возмущениями $h_{\mu\nu}$. 

Выражение для обратной метрики дано лишь в виде формального ряда:
\begin{align}\label{inverse_metric_with_perturbations}
  g^{\mu\nu}=\bar{g}^{\mu\nu} + \sum\limits_{n=1}^\infty (-\kappa)^n (h^n)^{\mu\nu} = \bar{g}^{\mu\nu} - \kappa h^{\mu\nu} + \kappa^2 h^{\mu\sigma} h_\sigma^{~\nu} + \cdots~.
\end{align}
Следуя формальным правилам умножения матриц можно показать, что \eqref{inverse_metric_with_perturbations} действительно играет роль обратной метрики:
\begin{align}
  \left(\bar{g}_{\mu\sigma} + \kappa h_{\mu\sigma}\right) \left[\bar{g}^{\sigma\nu} +\sum\limits_{n=1}^\infty (-\kappa)^n (h^n)^{\mu\nu}\right]=\delta_\mu^\nu~.
\end{align}
Тем не менее, для того, чтобы выражение \eqref{inverse_metric_with_perturbations} сходилось как степенной ряд необходимо, чтобы возмущения $h_{\mu\nu}$ были достаточно малы относительно $\kappa^{-1}$. На физическом уровне это соответствует тому, что мы рассматриваем слабо возмущённое гравитационное поле, чьи возмущения малыми по сравнению с массой Планка. В полном согласии с логикой, приведённой в разделе \ref{the_overview}, этот формализм даёт пертурбативное описание слабого гравитационного поля, но не применим для сильных полей.

Используя формулу \eqref{metric_with_perturbations} можно получить выражения для всех геометрических величин в любом порядке по $h_{\mu\nu}$, набор нужных формул приведён в приложении \ref{appendix_metric_perturbations}. Мы используем обозначения из работы \cite{tHooft:1974toh}. Фоновые величины мы будем помечать чертой над символом. Величины первого порядка по $h_{\mu\nu}$ мы будем обозначать одинарным подчёркиванием; величины второго порядка -- двойным подчёркиванием. Таким образом, в наших обозначениях $\bar{g}_{\mu\nu}$ есть фоновая метрика; $\underline{g}^{\mu\nu}= - \kappa h^{\mu\nu}$ если часть $g^{\mu\nu}$ линейная по $h_{\mu\nu}$; а также $\uunderline{g}^{\mu\nu}=\kappa^2 h^{\mu\sigma} h_{\sigma}^{~\nu}$. Все манипуляции с индексами производятся при помощи фоновой метрики $\bar{g}_{\mu\nu}$.

Физический смысл этих действий состоит в следующем. В традиционной формулировке ОТО динамической переменной, описывающими гравитационное поля, служит метрика $g_{\mu\nu}$. В данном подходе динамической переменной служит матрица возмущений фонового пространства-времени $h_{\mu\nu}$. Строго говоря, этот формализм представляет ОТО как теорию поля симметричного тензора второго ранга над искривлённым пространством-временем с метрикой $\bar{g}_{\mu\nu}$. Тем самым мы перешли от геометрического описания, данного в терминах искривлённой геометрии, к полевому, данному в терминах физических полей над фоновым пространством-временем. Такой переход оправдан и не противоречит базовым постулатам ОТО, так как оба способа описания, следуя Эйнштейну, эквивалентны, то есть описывают одну и ту же физическую картину мира разными способами. Однако, в отличии от геометрического описания, полевое описание легче поддаётся квантованию.

При помощи этого формализма можно получить описание динамики малых метрических возмущений в любой модели гравитации. В этом разделе мы ограничимся ОТО как наиболее релевантной теорией. Её действие имеется следующие компоненты:
\begin{align}
  \mathcal{A}[g_{\mu\nu}]\overset{\text{def}}{=} & -\cfrac{1}{16\pi G} \int d^4x \sqrt{-g} (R-2\Lambda)~, \\
  \underline{\mathcal{A}}=&\cfrac{\kappa}{16\pi G} \int d^4x \sqrt{-\bar{g}} \left[ \bar{R}_{\mu\nu} -\cfrac12 \bar{R} ~\bar{g}_{\mu\nu} + \Lambda ~\bar{g}_{\mu\nu} \right] h^{\mu\nu}~, \label{GR_linear_part} \\
  \uunderline{\mathcal{A}} =&-\cfrac{\kappa^2}{16\pi G} \int d^4 x\sqrt{-\bar{g}} \left[\cfrac14 h^{\mu\nu} \square h_{\mu\nu} -\cfrac14 h\square h +\cfrac12 h^{\mu\nu} \nabla_{\mu\nu} h -\cfrac12 h^{\mu\nu} \nabla^\sigma{}_\mu h_{\nu\sigma}\right. \nonumber \\
    &\left. -\cfrac14 \left( h^2_{\mu\nu} - \cfrac12 ~ h^2 \right) (\bar{R} -2 \Lambda) + (h^{\mu\sigma} h_\sigma{}^\nu - \cfrac12 h ~h^{\mu\nu}) \bar{R}_{\mu\nu} \right] ~. \label{GR_quadratic_part}
\end{align}
Фоновый член $\bar{S}$ не зависит от возмущений и включён в бесконечный нормирующий множитель. Линейный по $h_{\mu\nu}$ член \eqref{GR_linear_part} играет роль члена взаимодействия с линейным внешним источником:
\begin{align}
  \bar{J}_{\mu\nu}[\bar{g}] =\cfrac{2}{\kappa}\left[ \bar{R}_{\mu\nu} -\cfrac12 \bar{R} \bar{g}_{\mu\nu} +\Lambda \bar{g}_{\mu\nu} \right]. \label{background_perturbation_source}
\end{align}
Квадратичный по возмущениям член $\uunderline{S}$ описывает собственную динамику возмущений и играет роль кинетического члена; наконец, члены $O(h^3_{\mu\nu})$ описывают взаимодействие нескольких гравитонов. Разложение действия ОТО до третьего и четвёртого порядка по $h_{\mu\nu}$ приведены в работе \cite{Goroff:1985th}.

Теперь у нас имеется всё необходимое для квантования малых метрических возмущений над фоновым пространством-временем. Существование ненулевого космологического члена приводит к целому ряду специфических проблем, которые мы подробнее обсудим в следующем разделе \ref{quantum_GR_beyond_tree_level}. Поэтому до конца этого раздела мы ограничимся простым случаем $\Lambda=0$ и зафиксируем фоновую метрику на вакуумных уравнениях Эйнштейна. Это позволит нам взять в качестве фона плоское пространство-время и создать квантовую теорию малых возмущений на его фоне.

Удобно использовать набор дифференциальных операторов, определённых в работах Риверса и фон Ньюенхайзена \cite{Rivers1964,VanNieuwenhuizen:1973fi}, образующих полный базис проективных операторов. Мы приводим их импульсное представление и алгебру в приложении \ref{The_Operators}. Через операторы Риверса-фон Ньюенхайзен квадратичная часть действия для ОТО на плоском фоне \eqref{GR_quadratic_part} дана следующим выражением:
\begin{align}
  \uunderline{\mathcal{A}}=-\cfrac12 \int d^4x  ~ h^{\mu\nu}\left[P^2_{\mu\nu\alpha\beta} - 2 P^0_{\mu\nu\alpha\beta}\right] \square h^{\alpha\beta}~.
\end{align}
Генерирующий функционал же этой системы задан следующим выражением (с точностью до бесконечного нормирующего множителя):
\begin{align}
  \begin{split}
    Z&= \int\mathcal{D}[h] \exp\left[ i ~\mathcal{A}[\eta_{\mu\nu} + \kappa h_{\mu\nu}]\right] = \\
    &=\int\mathcal{D}[h] \exp\left[- \cfrac{i}{2} \int d^4 x  ~ h^{\mu\nu} ~\left[ P^2_{\mu\nu\alpha\beta} - 2P^0_{\mu\nu\alpha\beta} \right]~ \square h^{\alpha\beta} + O\left(h^3_{\mu\nu}\right)\right]~.
  \end{split}
\end{align}

Квадратичная часть действия является калибровочно-инвариантной и не может быть обращена. Эта ситуация стандартна для всех калибровочных моделей и она может быть разрешена с помощью техники духов Фаддеева-Попова \cite{Faddeev:1967fc,Faddeev:1973zb}. Мы не будем приводить подробное обсуждение этой техники, так как её применение в данном контексте полностью аналогично случаю калибровочных полей (более подробное обсуждение может быть найдено в работе \cite{Faddeev:1973zb}). Мы использует следующий член, фиксирующий фейнмановскую калибровку:
\begin{align}
  \mathcal{A}_\text{gf} = \int d^4 x  \left(\pd_\mu h^{\mu\nu}-\cfrac12 ~\pd^\nu h \right)^2 ~.
\end{align}
В этой калибровке духовые степени свободы полностью отделены от гравитационного сектора и могут быть включены в бесконечный нормирующий множитель. Генерирующий функционал же принимает следующий вид:
\begin{align}
  Z = \underset{\scriptsize{\begin{matrix}\text{физические} \\ \text{состояния}\end{matrix}}}{\int\limits\mathcal{D}[h]} \exp\left[  -\cfrac{i}{2} ~\int d^4x ~ h^{\mu\nu} ~\cfrac{C_{\mu\nu\alpha\beta}}{2}~\square h^{\alpha\beta} +i\, \mathcal{A}_\text{int} \right],
\end{align}
тут символы $C_{\mu\nu\alpha\beta}$ заданы следующим выражением:
\begin{align}
  C_{\mu\nu\alpha\beta}\overset{\text{def}}{=} \eta_{\mu\alpha} \eta_{\nu\beta} +\eta_{\mu\beta}\eta_{\nu\alpha}-\eta_{\mu\nu} \eta_{\alpha\beta}~.
\end{align}
Подчеркнём отдельно, что действие $\mathcal{A}_\text{int}$, описывающее взаимодействие гравитонов, имеет бесконечное число членов. Чтобы завершить квантование нужно ввести формальный внешний ток $J_{\mu\nu}$ в присутствии которого генерирующий функционал (в калибровке фейнмана) будет выглядеть следующим образом \cite{zinn-justin_path_2011,RN14}:
\begin{align}
  Z[J] &= \underset{\scriptsize{\begin{matrix}\text{физические} \\ \text{состояния}\end{matrix}}}{\int\mathcal{D}[h]} \exp\left[ -\cfrac{i}{2}\int d^4 x ~ h^{\mu\nu} \cfrac{C_{\mu\nu\alpha\beta}}{2}~\square h^{\alpha\beta} + i \mathcal{A}_\text{int}[h] + i~ J^{\mu\nu} h_{\mu\nu}\right] \nonumber \\
  &=\exp\left[i \mathcal{A}_\text{int}\left[\cfrac{1}{i} \cfrac{\delta}{\delta J} \right]\right] \exp\left[\cfrac{i}{2}~ J^{\mu\nu}~ ~ \cfrac{C_{\mu\nu\alpha\beta}}{2}~\square^{-1} J^{\alpha\beta}\right].\label{generating_functional_0}
\end{align}
Подробный алгоритм действий, фиксирующих калибровку для ОТО и квадратичных моделей гравитации, приведён в работе \cite{Accioly:2000nm}.

Генерирующий функционал \eqref{generating_functional_0} позволяет нам получать выражения для всех многочастичных корреляционных функций в любом порядке по теории возмущений. В частности, в древесном приближении пропагатор гравитона в импульсном представлении задан следующим выражением:
\begin{align}
  \langle 0 \lvert h_{\mu\nu} h_{\alpha\beta} \rvert 0 \rangle = \cfrac{1}{i} \cfrac{\delta}{\delta J^{\mu\nu}} \cfrac{1}{i} \cfrac{\delta}{\delta J^{\alpha\beta}} \left.\cfrac{Z[J]}{Z[0]}\right|_{J=0} = \cfrac{i}{2} \cfrac{C_{\mu\nu\alpha\beta}}{k^2} + O(\kappa^2)~.
\end{align}

Материальные степени свободы включаются в построенную квантовую теорию по полной аналогии с материальными степенями свободы в калибровочных теориях. В качестве простейшего примера можно привести генерирующий функционал для системы, состоящей из малых метрических возмущений (в фейнмановской калибровке) и скалярного поля:
\begin{align}\label{generating_functional_1}
  Z=\underset{\scriptsize{\begin{matrix}\text{физические} \\ \text{состояния}\end{matrix}}}{\int\underbrace{\mathcal{D}[h]}}\mathcal{D}[\phi] \exp & \left[\int d^4 x \left\{ -\cfrac{i}{2}~h^{\mu\nu} ~\cfrac{C_{\mu\nu\alpha\beta}}{2} ~\square h^{\alpha\beta}-\cfrac{i}{2} ~\phi(\square+m^2) \phi\right. \right. \\
    & \left. \left.-i\,\cfrac{\kappa}{4}\, h^{\mu\nu}\left(C_{\mu\nu\alpha\beta} \pd^\alpha \phi \pd^\beta \phi + \eta_{\mu\nu} m^2 \phi^2\right) +\cdots\right\} \right]. \nonumber
\end{align}
В данном выражении из сектора взаимодействия мы оставили лишь ведущий член взаимодействия гравитонов с тензором энергии-импульса скалярного поля.

Таким образом, вопреки распространённому убеждению, можно построить квантовую теорию малых метрических возмущений над плоским про\-стран\-ством\--вре\-ме\-нем в линейном приближении. Однако, как сказано выше, попытка расширить применение этого подхода на уровень первой радиационной поправки не приводит к удовлетворительным результатам. Как мы покажем в следующем разделе, существование размерной константы взаимодействия -- массы Планка -- приводит к тому, что теорию невозможно перенормировать стандартными методами. Вследствие чего стандартная программа квантования вряд ли может быть признана удовлетворительной.

\section{Проблемы пертурбативного квантования гравитации}\label{quantum_GR_beyond_tree_level}

Пертурбативный подход к квантованию гравитации имеет две основные проблемы. Во-первых, пертурбативная квантовая ОТО не может быть перенормирована даже на фоне плоского пространства-времени. Во-вторых, перенормируемые теории квантовой гравитации, построенные пертурбативным методом, содержат духовые состояния. Отдельную проблему представляет из себя космологическая постоянная.

Неперенормируемость квантовой ОТО на уровне первой радиационной поправки была показана в работе \cite{tHooft:1974toh}. В отсутствии материи квантовая ОТО может быть перенормируема только на массовой оболочке. При включении в модель матери перенормируемость теряется даже на уровне массовой оболочки. Окончательно вопрос о неперенормируемости ОТО был разрешён в \cite{Goroff:1985th}, где было показано, что на уровне двух петель ОТО неперенормируема даже на уровне массовой оболочки (смотри также \cite{vandeVen:1991gw,Barvinsky:1987uw}).

Показать неперенормируемость квантовой пертурбативной ОТО можно следующими прямым вычислениями. Для простоты ограничимся системой из гравитационного поля и безмассового скаляра. Правила Фейнмана на древесном уровне можно получить непосредственно по генерирующему функционалу \eqref{generating_functional_1} (в калибровке Фейнмана):
\begin{align}
  \begin{gathered}
    \begin{fmffile}{pic01}
      \begin{fmfgraph}(40,40) 
        \fmfleft{i}
        \fmfright{o}
        \fmf{dbl_wiggly}{i,o}
      \end{fmfgraph}
    \end{fmffile}
  \end{gathered}
  &= \cfrac{i}{2} \cfrac{C_{\mu\nu\alpha\beta}}{k^2} ~, &
  \begin{gathered}
    \begin{fmffile}{pic02}
      \begin{fmfgraph}(40,40)
        \fmfleft{i}
        \fmfright{o}
        \fmf{dashes}{i,o}
      \end{fmfgraph}
    \end{fmffile}
  \end{gathered}
  &= ~\cfrac{i}{k^2} ~,
\end{align}
\begin{align}
  \begin{gathered}
    \begin{fmffile}{pic03}
      \begin{fmfgraph*}(60,60)
        \fmfleft{i1,i2}
        \fmfright{o}
        \fmf{dashes_arrow,label=$p$}{i1,v}
        \fmf{dashes_arrow,label=$q$,label.side=right}{v,i2}
        \fmf{dbl_wiggly,tension=2}{v,o}
        \fmfdot{v}
        \fmflabel{$\mu\nu$}{o}
      \end{fmfgraph*}
    \end{fmffile}
  \end{gathered} \hspace{.7cm}
  &= -i\,\cfrac{\kappa}{4}\, C_{\mu\nu\alpha\beta}\, p^\alpha\, q^\beta~. \nonumber
\end{align}
С их помощью можно вычислить первую радиационную поправку к собственной энергии гравитона (определение символов $\Theta_{\mu\nu}$ приведено в приложении \ref{The_Operators}):
\begin{align}
  \begin{gathered}
    \begin{fmffile}{pic04}
      \begin{fmfgraph*}(60,60)
        \fmfleft{i}
        \fmfright{o}
        \fmf{phantom}{i,vl,vr,o}
        \fmfdot{vl,vr}
        \fmffreeze
        \fmf{dashes,left}{vl,vr}
        \fmf{dashes,left}{vr,vl}
        \fmflabel{$\mu\nu$}{vl}
        \fmflabel{$\alpha\beta$}{vr}
      \end{fmfgraph*}
    \end{fmffile}
  \end{gathered}
  & = i \Pi_{\mu\nu\alpha\beta}(q^2) \\
  & =-\cfrac{i ~\kappa^2 q^4}{7670\pi^2} \left[\Theta_{\mu\alpha}\Theta_{\nu\beta} +\Theta_{\mu\beta} \Theta_{\nu\alpha} +6 \Theta_{\mu\nu} \Theta_{\alpha\beta}\right] \cfrac{1}{d-4} + O(1)~. \nonumber
\end{align}
Лагранжиан контрчленов, требуемых для перенормировки этой диаграммы, выражается через операторы, генерируемые вкладами квадратичными по кривизне:
\begin{align}
  \Delta \mathcal{L}=h^{\mu\nu} \Pi_{\mu\nu\alpha\beta} h^{\alpha\beta} =\cfrac{1}{960\pi^2} \left( R_{\mu\nu}^2 +\cfrac12 R^2 \right) \cfrac{1}{d-4} + O(1)~.
\end{align}
Так как мы работали с моделью, содержащей материю, ни тензор Риччи, ни скалярная кривизна не обращаются в нуль на массовой оболочке (на уравнениях поля), вследствие чего лагранжиан контрчленов не обращается в нуль. Невозможность компенсировать бесконечный вклад контрчленов, связанных с собственной энергией гравитона, и означает невозможность перенормировки теории стандартным методом.

Как уже было отмечено, впервые проблемы с перенормируемостью квантовой версии ОТО были найдены в работе 'т Хоофта и Вельтмана \cite{tHooft:1974toh}. Они показали, что в рамках квантовой версии ОТО в отсутствии материи первая петлевая поправка к собственной энергии электрона также требует членов квадратичных по кривизне для перенормировки:
\begin{align}
  \begin{gathered}
    \begin{fmffile}{pic05}
      \begin{fmfgraph}(50,50)
        \fmfleft{i}
        \fmfright{o}
        \fmf{dbl_wiggly,tension=2}{i,l}
        \fmf{dbl_wiggly,left=1,tension=.5}{l,r,l}
        \fmf{dbl_wiggly,tension=2}{r,o}
      \end{fmfgraph}
    \end{fmffile}
  \end{gathered}
  \to \Delta \mathcal{L}\sim\left[\cfrac{1}{72}R^2+\cfrac{1}{60} \left(R_{\mu\nu}^2-\cfrac13 R^2\right) \right]\cfrac{1}{d-4} + O(1)~.
\end{align}
В рамках ОТО на массовой оболочке (на вакуумных решениях уравнений Эйнштейна) тензор Риччи обращается в нуль, вследствие чего требуемый контрчлен тождественно равен нулю и не требует перенормировки. Однако, на уровне второй петлевой поправки, как было показано Горовым и Сагнотти \cite{Goroff:1985th}, подобное явление не реализуется. Для перенормировки второй петлевой поправки требуются члены с высшими производными вида $R \square R$, которые не обращаются в нуль даже на решениях вакуумных уравнений Эйнштейна. Таким образом, квантовую пертурбативную версию ОТО нельзя перенормировать стандартными методами.

Более того, можно показать, что вклад гравитационного взаимодействия в собственную энергию любой частицы может требовать контрчлены, содержащие высшие производные. Исходя из соображений размерности структуру вклада можно качественно восстановить исходя из того, что гравитационное взаимодействие квадратично по импульсам и трёхчастичное взаимодействие подавлено фактором $\kappa$. При больших значениях импульса, переносимого виртуальным гравитоном, эта поправка будет иметь квадратичную по импульсу расходимость:
\begin{align}
  \begin{gathered}
    \begin{fmffile}{pic06}
      \begin{fmfgraph*}(30,30)
        \fmfleft{i}
        \fmfright{o}
        \fmf{dbl_wiggly,left=1}{i,o,i}
        \fmfdot{i,o}
      \end{fmfgraph*}
    \end{fmffile}
  \end{gathered}
  &~\simeq \kappa^2 ~\int \cfrac{d^4 k}{(2\pi)^4} \cfrac{p^2 (p-k)^2}{k^2 (p-k)^2} \underset{k\to\infty}{\to} p^2 \kappa^2 \int k dk ~.
\end{align}
В силу размерности этого вклада, лагранжиан требуемых контрчленов должен быть пропорциональным $\kappa^2\square^2$. С формальной точки зрения неперенормируемость возникает из-за того, что взаимодействие с гравитонами квадратично по импульсам. Для того, чтобы такой член взаимодействия имел корректную размерность необходимо вводить размерную постоянную $\kappa$. Вследствие чего именно импульсная структура взаимодействия генерирует операторы с высшими производными на уровне радиационных поправок.

Вторая проблема, указанная выше, возникает при попытке улучшить перенормируемость теории, включив в микроскопический лагранжиан члены с высшими производными, необходимые для перенормировки:
\begin{align}\label{action_s_2}
  \mathcal{A}_\text{GR} \to \mathcal{A}_\text{Stelle}=\int d^4 x \sqrt{-g} \left[-\cfrac{1}{16\pi G} \, R + c_1 R^2 + c_2 R_{\mu\nu}^2  \right]~.
\end{align}
В данном выражении мы считаем $c_1$ и $c_2$ произвольными постоянными. Заметим, что построенная таким образом модель будет отличаться от ОТО уже на древесном уровне, что поднимает вопрос об её интерпретации.

Пертурбативное квантование модели \eqref{action_s_2} было проведено работах Штэлле \cite{Stelle:1977ry,Stelle:1976gc}, где было показано, что модель является перенормируемой, что разрешает часть проблем квантовой ОТО. Но в то же время спектр состояний модели содержит духовые степени свободы. Даже в древесном приближении пропагатор метрических возмущений, описываемых действием \eqref{action_s_2}, имеет следующую калибровочно-инвариантную часть (вывод обсуждается в приложении \ref{quadratic_gravity_propagator} и в работе \cite{Accioly:2000nm}):
\begin{align}
  &-i \langle 0 \rvert h_{\mu\nu} h_{\alpha\beta} \lvert 0 \rangle = \\
  &=\cfrac{1}{k^2} \left[P^2_{\mu\nu\alpha\beta} -\cfrac12 P^0_{\mu\nu\alpha\beta}\right] - \cfrac{P^2_{\mu\nu\alpha\beta}}{k^2-m_2^2} + \cfrac{P^0_{\mu\nu\alpha\beta}}{k^2-m_0^2} + O(\kappa^2)~. \nonumber
\end{align}
Размерные константы $m_0$ и $m_2$ определены через $c_1$ и $c_2$ следующим образом:
\begin{align}\label{gravity_eft_massess}
  \begin{cases}
    m_2^2 &= - \cfrac{2}{c_2 \kappa^2}~,  \\
    m_0^2 &=\cfrac{1}{\kappa^2(3c_1-c_2)} ~.
  \end{cases}
\end{align}
Пропагатор имеет новые полюса, которые описывают новые степени свободы. Таким образом, модель \eqref{action_s_2} описывает стандартный гравитон (безмассовые возмущения со спином $2$), скаляры с массой $m_0$ и духи со спином $2$ и массой $m_2$. Данный результат может быть также получен и непертурбативными методами \cite{Hindawi:1995an,Barth:1983hb}. Подчеркнём особо, что данные духовые состояния никак не связаны ни с калибровочной инвариантностью, ни с духами Фаддеева-Попова. Как следствие, эти состояния должны рассматриваться как реальные возбуждения системы, переносящие отрицательную кинетическую энергию, что делает модель неприменимой.

Для полноты изложения необходимо заметить, что есть возможность согласованно исключить духовые состояния из теории. Прежде всего, заметим, что не все члены микроскопического действия дают вклад в пропагатор возмущений на фоне плоского пространства-времени. Члены, содержащие третью и более высокие степени тензора Римана, не содержат членов $O\left(h_{\mu\nu}^2\right)$ \cite{Calmet:2019odl}. Как следствие, в самом общем случае для произвольной теории гравитации микроскопическое действие, определяющее спектр возмущений теории над плоским пространством-временем, имеет следующий вид:
\begin{align}\label{action_IDG}
  &\mathcal{A}_\text{quadratic}= \\
  &=\int d^4 x \sqrt{-g} \left[-\cfrac{1}{16\pi G}\, R + R\,f_1(\square) R + R_{\mu\nu} f_2(\square) R^{\mu\nu} + R_{\mu\nu\alpha\beta}\,f_3(\square)\,R^{\mu\nu\alpha\beta} \right] ~. \nonumber
\end{align}
Действия такого вида активно изучаются в рамках моделей гравитации с бесконечным числом производных \cite{Kuzmin:1989sp,Tomboulis:2015esa,Modesto:2012ys,Modesto:2017sdr,Modesto:2014lga,Briscese:2018oyx,Briscese:2018bny,Briscese:2019rii,Biswas:2011ar,Krasnikov:1987yj,Biswas:2013cha}.

Если функции $f_1$, $f_2$ и $f_3$ подобраны особым образом, то действие \eqref{action_IDG} может не содержать духовых степеней свободы \cite{Modesto:2017sdr,Tomboulis:2015esa,Biswas:2013kla,Biswas:2011ar}, что разрешает проблему духов с формальной точки зрения. Тем не менее, нет ни оснований полагать, что микроскопическое действие для гравитации должно иметь вид \eqref{action_IDG}, ни физических соображений, способных однозначно зафиксировать вид функций $f_i$. Таким образом, на данный момент нет оснований полагать, что подобный сценарий реализуется в природе.

Стоит также отметить, что проблема духов стала актуальна в последние годы. Духовые степени свободы появляются в моделях с высшими производными естественным образом из-за неустойчивости Остроградского \cite{Ostrogradsky:1850fid,Woodard:2006nt}. Долгое время считалось, что системы с духовыми степенями свободы имеют состояния с отрицательной вероятностью. Однако, в 2002 году Хокингом и Хертогом была найдена система с высшими производным, имеющая не унитарную эволюцию, но и не содержащая ни состояний с отрицательной нормой, ни отрицательных вероятностей перехода \cite{Hawking:2001yt}. Вслед за этим было показано, что отрицательные вероятность можно исключить из модели осциллятора с высшими производными (осциллятор Пэйса-Улембека) при помощи $PT$-симметрии \cite{Bender:2007wu}. Наконец, работы \cite{Smilga:2008pr,Smilga:2005gb,Mannheim:2004qz} подняли вопрос о возможности согласования унитарной эволюции с духовыми степенями свободы. На данный момент консенсус относительно возможности унитарного описания духов не достигнут. Тем не менее, даже если такое описание возможно, то модели с  духовыми степенями свободы также будут иметь критическую неустойчивость и не могут быть признаны реалистичными \cite{Woodard:2006nt}. Более подробно проблема духов обсуждается в работах \cite{Woodard:2006nt,Sbisa:2014pzo,Salles:2018ccb}.

Завершая обсуждение проблем квантовой ОТО особо стоит отметить роль космологической постоянной в пертурбативной квантовой ОТО. Космологическая постоянная является свободным параметром теории \cite{Lovelock:1971yv,Berti:2015itd} и в самом общем случае нет оснований считать её нулевой. Эмирические данные явно указывают на то, что её величина отлична от нуля \cite{Aghanim:2018eyx}. Сама по себе проблема космологической постоянной обсуждается во множестве работ, например в \cite{Zeldovich:1968ehl,Weinberg:1988cp,Padilla:2015aaa}; здесь мы лишь кратко обсудим её суть.

Введение в микроскопическое действие ненулевой космологической постоянной существенно затрудняет применение программы пертурбативного квантования. Следуя обозначенной ранее логике, космологическую постоянную, существующую на уровне микроскопического действия, следует рассматривать как внешний источник гравитационного поля. Однако, этот источник не локализован в пространстве и, как следствие, генерирует новые возмущения в каждой точке пространства-времени. Из-за этого даже сколь угодно малую космологическую постоянную нельзя считать пертурбативным возмущением, однако, можно рассматривать малые возмущения на макроскопическом фоне, определённом космологической постоянной.

Фоновое пространство-время, в таком случае, будет иметь постоянную кривизну, определяемую величиной космологической постоянной. Если про\-стран\-ство\--вре\-мя имеет отрицательную кривизну, то оно описывается пространство анти-де Ситтера, которое не является глобально-гиперболичным. Это означает, что задачи Коши не может быть однозначно решена на всём пространстве-времени \cite{RN20,RN18}, а теория теряет предсказательную способность. Если пространство-время имеет положительную кривизну, то это пространство де Ситтера, в котором невозможно ввести $S$-матрицу \cite{Witten:2001kn,Bousso:2004tv}. Связано это, от части, с невозможностью введения асимптотических состояний. В плоском пространстве-времени всегда возможно выбрать состояния, имеющие хорошо определённый импульс на пространственной бесконечности. Пространство де Ситтера, в свою очередь, описывает ускоренно расширяющуюся вселенную. В такой вселенной при удалении на бесконечность энергия любого состояния будет неограниченно расти. Это обстоятельство и препятствует введению базисных состояний и $S$-матрицы.

Наконец, в рамках квантовой теории возникает проблема, впервые найденная в работе Якова Борисовича Зельдовича \cite{Zeldovich:1968ehl}. Величина $\Lambda$, точно также и величины всех других констант, испытывает влияние квантовых поправок и, как следствие, начинает зависеть от энергии измерения. Формула для величины $\Lambda$ дана следующим рядом:
\begin{align}
  \begin{gathered}
    \begin{fmffile}{pic07}
      \begin{fmfgraph}(40,40)
        \fmfleft{l}
        \fmfright{r}
        \fmftop{t}
        \fmfbottom{b}
        \fmf{phantom}{l,v,r}
        \fmf{phantom}{b,v,t}
        \fmffreeze
        \fmfblob{10}{v}
      \end{fmfgraph}
    \end{fmffile}
  \end{gathered}
  + ~
  \begin{gathered}
    \begin{fmffile}{pic08}
      \begin{fmfgraph}(40,40)
        \fmfleft{l}
        \fmfright{r}
        \fmftop{t}
        \fmfbottom{b}
        \fmf{phantom}{l,v,r}
        \fmf{phantom}{b,v,t}
        \fmffreeze
        \fmfblob{10}{v}
        \fmf{dbl_wiggly}{l,v}
      \end{fmfgraph}
    \end{fmffile}
  \end{gathered}
  + ~
  \begin{gathered}
    \begin{fmffile}{pic09}
      \begin{fmfgraph}(40,40)
        \fmfleft{l}
        \fmfright{r}
        \fmftop{t}
        \fmfbottom{b}
        \fmf{phantom}{l,v,r}
        \fmf{phantom}{b,v,t}
        \fmffreeze
        \fmfblob{10}{v}
        \fmf{dbl_wiggly}{l,v,r}
      \end{fmfgraph}
    \end{fmffile}
  \end{gathered}
  ~ + ~
  \begin{gathered}
    \begin{fmffile}{pic10}
      \begin{fmfgraph}(40,40)
        \fmfleft{l}
        \fmfright{r}
        \fmftop{t}
        \fmfbottom{b}
        \fmf{phantom}{l,v,r}
        \fmf{phantom}{b,v,t}
        \fmffreeze
        \fmfblob{10}{v}
        \fmf{dbl_wiggly}{l,v,r}
        \fmf{dbl_wiggly}{v,t}
      \end{fmfgraph}
    \end{fmffile}
  \end{gathered}
  ~ + ~ \cdots
\end{align}
Соображения размерности и прямые вычисления \cite{Zeldovich:1968ehl,Weinberg:1988cp} показывают, что это выражение имеет следующий вид \cite{Padilla:2015aaa}:
\begin{align}
  \Lambda_\text{1-loop} \sim m^4 \left[ \cfrac{1}{\varepsilon} +\text{finite part} - \ln\left(\cfrac{m^2}{\mu^2}\right) \right] ~.
\end{align}
Тут $m$ обозначает массу наиболее тяжёлой частицы, присутствующей в модели. Первое затруднение возникает в связи с тем, что величина конечной части выражения определяется четвёртой степенью массы $m$. Реалистичные оценки величины $m^4$, сделанные при использовании массы $t$-кварка и электрона, на сотни порядков отличаются от наблюдаемой величины $\Lambda$-члена. Вследствие этого процедура вычитания должна быть проведена с невероятной точностью, что часто называется проблемой тонкой настройки модели. Во-вторых, измерения величины $\Lambda$-члена и масс элементарных частиц производятся на различным энергетических масштабах. Величина космологического члена измерена в современной Вселенной, то есть при крайне малых энергиях, в то время как величины масс элементарных частиц измеряются в режиме высоких энергий. Иными словами, мы можем вычислить величину $\Lambda$-члена в той области, которую не может измерить и наоборот.

Завершая раздел, стоит суммировать базовые проблемы программы пертурбативного квантования гравитации. Во-первых, радиационные поправки к квантовой ОТО требуют контрчленов, отсутствующих в исходном микроскопическом действии. Как следствие, пептурбативная квантовая ОТО не может быть перенормирована стандартными методами квантовой теории. Во-вторых, для перенормировки радиационных поправок требуются члены, содержащие высшие производные. Включение таких членов в микроскопическое действие приведёт к возникновению духовых степеней свободы. В-третьих, программа пертурбативного квантования испытывает проблемы с ненулевой космологической постоянной. В зависимости от её величины, теория или испытывает проблемы с предсказательной способностью, или делает невозможным использование $S$-матрицы. Наконец, влияние квантовых эффектов на величину космологической постоянной требует точной настройки параметров вычитания. По совокупности этих причин программа пертурбативного квантования гравитации в современном виде не может быть признана удовлетворительной.

\section{Каноническая квантовая гравитация}\label{nonperturbative_quantum_gravity}

Создание канонической квантовой теории гравитации следует программе канонического квантования, изложенной в работе Дирака \cite{DiracPaulAdrienMaurice1958TPoQ}. Квантовое описание системы даётся при помощи операторов, которые в этом разделе мы обозначаем циркумфлексом $\hat{\cdot}$, и состояний $\ket{\psi}$. Операторы подчинены коммутационным соотношениям $[\cdot,\cdot]$ и образуют замкнутую алгебру. Вся динамика системы описывается оператором эволюции $\widehat{S}(r,t')$, который переводит состояние системы в начальный момент времени $\ket{\psi(t')}$ в состояние в конечный момент времени $\ket{\psi(t)}$:
\begin{align}
  \widehat{S}(t,t') \ket{\psi(t')}=\ket{\psi(t)} ~.
\end{align}

Метод канонического квантования сам по себе не лишён трудностей, что было отмечено на самом раннем этапе его развития \cite{DiracPaulAdrienMaurice1958TPoQ}. Главной его проблемой является невозможность однозначного построения квантовой теории по классической \cite{DiracPaulAdrienMaurice1958TPoQ,RN16,Sardanashvili2017}. Эта проблема имеет два аспекта.

Во-первых, существует проблема упорядочения операторов, так как по классическим выражениям для физических величин невозможно однозначно определить порядок квантовых операторов, используемых в квантовой теории. Эта проблема хорошо раскрыта в классических учебниках \cite{RN14,RN4,Bogolubov} и мы не будем обсуждать её подробно.

Во-вторых, по классическим теориям с калибровочной симметрией, к которым относится и гравитация, невозможно однозначно восстановить базис физических состояний. Проиллюстрировать эту проблему проще всего на примере электромагнитного поля. Электромагнитное поле имеет калибровочную симметрию, которой мы сопоставим оператор $\widehat{\mathfrak{G}}$. Рассмотрим два собственных состояния оператора электромагнитного поля $\widehat{\mathcal{A}}_\mu$ -- состояния $\ket{A_\mu}$ и $\ket{A'_\mu}$. Если эти собственные состояния связаны калибровочной симметрией $\mathfrak{G}\ket{A_\mu}=\ket{A'_\mu}$, то, следуя логике классической теории, мы должны ожидать, что все наблюдаемые величины на состояниях $\ket{A_\mu}$ и $\ket{A'_\mu}$ должны совпадать. Однако, это и означает, что состояния $\ket{A_\mu}$ и $\ket{A'_\mu}$ описывают одно и тоже физическое состояние системы, которому формально приписали разные значения $A_\mu$. Иными словами, базис собственных состояний оператора поля $\widehat{\mathcal{A}}_\mu$ переопределен в том смысле, что одному физическому состоянию невозможно однозначно приписать потенциал поля $A_\mu$. В канонической квантовой теории калибровочных полей эта проблема разрешается сужением базиса при помощи квантовых калибровочных условий \cite{DiracPaulAdrienMaurice1958TPoQ}. Мы выбираем калибровочные условия и исключаем из рассмотрения все состояния, которые им не удовлетворяют. Например, в случае электродинамики мы можем выбрать калибровочное условие Лоренца $\pd_\mu A^\mu=0$ и рассматривать лишь состояния, удовлетворяющие квантовым условиям Лоренца:
\begin{align}
  \pd_\mu \hat{\mathcal{A}}^\mu \ket{\Psi}=0~.
\end{align}

Аналогичная ситуация имеет место и в случае гравитации. В качестве потенциала поля выступает метрика $g_{\mu\nu}$, в роли же напряжённостей выступают или символы Кристофеля $\Gamma^{\alpha}_{\mu\nu}$, если речь идёт о геодезическом движении, или тензор Римана, если речь идёт о полевых моделях. Роль калибровочной симметрии играет замена координат в пространстве-времени и выбор локального базиса. На уровне метрики калибровочная симметрия выражается следующими преобразованиями:
\begin{align}
  g_{\mu\nu} \to g_{\mu\nu} + \pd_\mu \zeta_\nu + \pd_\nu \zeta_\mu~,
\end{align}
где $\zeta_\mu$ -- произвольное векторное поле. Из-за этих симметрий метрика про\-стран\-ства\--времени не может быть определена однозначно, поэтому базис состояний, построенных на собственных значениях оператора метрики, также будет переопределен.

При попытке построения канонической квантовой теории гравитации к этим проблемам добавляется проблема времени. При изучении систем без учёта гравитации время выступает в качестве внешнего параметра в том смысле, что никакие физические процессы, протекающие внутри системы, не оказывают на него влияния. В случае с гравитационным полем (согласно нашим нынешним представлениям) установить внешнюю временную параметризацию принципиально невозможно \cite{Torre:1992rg}. Поэтому для корректной канонической теории квантовой гравитации необходимо заново вводить представление о времени. Проблема времени в канонической квантовой гравитации заслуживает отдельного обсуждения, лежащего за пределами этого обзора. Подробные обзоры проблемы времени можно найти в работах \cite{kuchar_time_2011,isham_canonical_1993,anderson_problem_2012}.

В этом разделе мы будем использовать консервативный подход к проблеме времени, называемый введением времени до квантования. А именно, в рамках классической теории мы введём произвольную временную параметризацию, после чего обратимся к исследованию алгебры Пуассона, генерируемой классической теорией. Основной проблемой данного подхода является создаваемое им спонтанное нарушение симметрии пространства-времени. Как отмечено выше, в рамках ОТО все возможные параметризации времени являются эквивалентными. Выбирая одну конкретную параметризацию мы спонтанно нарушаем симметрию между ними. Построенная таким образом квантовая теория будет иметь спонтанно выделенное направление времени. Основным недостатком такого подхода является то, что сам механизм спонтанного нарушения симметрии является внешним по отношению к модели и не возникает естественным образом.

Следует заметить, что от полной квантовой теории гравитации следует ожидать наличия механизма спонтанного нарушения симметрии между пространством и временем. Эмпирическим фактом, указывающим на необходимость его существования, является существование выделенной системы отсчёта в наблюдаемой Вселенной. Система отсчёта, сопутствующая материи при космологическом расширении, является спонтанно выделенной среди прочих, так как в ней вся материя Вселенной как целое покоится. Более того, как было отмечено ещё самим Эйнштейном \cite{Einstein:1917ce}, в расширяющейся Вселенной время можно связать с наблюдаемой величиной -- масштабным фактором -- в результате чего время перестанет играть роль произвольной координаты и станет физической величиной. Современные исследования предоставляют конкретные методы введения параметризационно-инвариантного времени в космологических моделях \cite{Pervushin:2011gz,Pervushin:2017zfj,Arbuzov:2019crw,Arbuzov:2017sfw}. Эти соображения предоставляют основания считать, что полная квантовая теория гравитации будет содержать конкретный механизм спонтанного нарушения симметрия между пространством и временем, ответственный за космологическое расширение.

Несмотря на то, что подход введения времени до квантования не разрешает проблему времени в целом, он позволяет получить конкретные математические выражения, необходимые для применения программы канонического квантования. Построенное таким образом описание гравитационного поля должно быть дано в терминах динамической геометрии трёхмерного пространства. Метод разбиения четырёхмерного пространства-времени на серию трёхмерных поверхностей был представлен в работе Арновитта, Дезера и Мизнера \cite{Arnowitt:1962hi} (стоит отметить также работы Дирака \cite{Dirac:1958jc,Dirac:1958sc}). Здесь мы приведём лишь краткое описание метода $3+1$ разбиения; детальные математические выкладки могут быть найдены в \cite{RN19}; необходимые математические теоремы о кривизне подмногообразия приведены в \cite{RN10}.

Прежде всего, мы введём координаты на пространстве-времени. Временную координату мы будем обозначать переменной $t$ и выберем координаты так, чтобы любая поверхность $\Sigma_t$ с фиксированным $t$ была поверхностью Коши. Иными словами, мы считаем, что задача Кошки может быть корректно поставлена для любого момента времени. Параметризация времени позволяет определить времени-подобный вектор $\tau^\mu$, задаваемый по временной переменной $t$ следующим образом:
\begin{align}
  \tau^\mu\nabla_\mu t =1~.
\end{align}
Таким образом, мы построили семейство поверхностей $\Sigma_t$, параметризуемых временем $t$, и времени-подобное векторное поле $\tau^\mu$.

Векторное поле $\tau^\mu$ нужно разбить на две компоненты. Первая компонента называется функцией хода $N$ и равняется проекции векторного поля $\tau^\mu$ на вектор нормальный к поверхностям $\Sigma_t$. Вторая компонента поля $\tau^\mu$ называется вектором сдвига $N_i$, и она определяется как компонента поля $\tau^\mu$ касательная к поверхностям $\Sigma_\tau$. Иными словами, вектор сдвига $N_i$ описывает собственную деформацию пространственных слоёв, а функция хода -- относительный сдвиг слоёв с течением времени.

Геометрические свойства каждой поверхности $\Sigma_t$ могут быть разделены на два класса -- внешнюю и собственную геометрии (extrinsic and intrinsic geometries). К собственной геометрии относят те свойства, которые не зависят от конкретного способа её вложения в четырёхмерное пространство; ко внешней же геометрии, напротив, относят лишь те свойства, что определяются конкретным способом вложения поверхности в пространство большей размерности. На практическом уровне свойства собственной геометрии описываются трёхмерным тензором кривизны $R^{(3)}_{ij}{}^a{}_b$. Следуя правилам дифференциальной геометрии, трёхмерный тензор кривизны определяет свойства параллельного переноса векторов, фиксированных на трёхмерной поверхности, по этой поверхности. Свойства внешней геометрии описываются тензором внешней кривизны $K_{ij}$. Этот тензор описывает отклонение вектора $\tau^\mu$ от нормали при параллельном переносе вдоль трёхмерной поверхности \cite{RN10}.

Таким образом, четырёхмерную метрику $g_{\mu\nu}$ теперь следует выразить через метрику трёхмерных слоёв $\gamma_{ij}$, функцию хода $N$ и вектор сдвига $N_i$. Эти десять независимых величин определяют собственную и внешнюю геометрии пространственных слоёв, причём тензор трёхмерной кривизны выражается только через трёхмерную метрику. Иными словами, вся информация о вложении трёхмерных поверхностей в четырёхмерное пространство содержится в функции хода и векторе сдвига.

В терминах внешней $K_{ij}$ и собственной кривизны $R^{(3)}$ действие для ОТО записывается в следующем виде \cite{RN19}:
\begin{align}\label{GR_ADM_action}
  \begin{split}
    S &= \cfrac{1}{16\pi G} \int d^4 x ~ \sqrt{-g} R \\
    &=\cfrac{1}{16\pi G} \int dt ~ \int d^3x ~N\left[ \sqrt{\gamma} R^{(3)} + K^{mn} G_{mnab} K^{ab}\right]~.
  \end{split}
\end{align}
В этом выражении $G_{mnab}$ обозначает метрику ДеВитта
\begin{align}
  \begin{split}
    G_{mnab}=\cfrac{1}{2\sqrt{\gamma}} \left(\gamma_{ma} \gamma_{nb}+\gamma_{mb}\gamma_{na}-\gamma_{mn}\gamma_{ab}\right) ~, \\
    G^{mnab}=\cfrac{\sqrt{\gamma}}{2} \left(\gamma^{ma}\gamma^{nb}+\gamma^{mb}\gamma^{na}-2\gamma^{mn}\gamma^{ab}\right) ~;
  \end{split}
\end{align}
внешняя кривизна дана следующей формулой:
\begin{align}
  K_{ij}= \cfrac{1}{2N}\left(\dot{\gamma}_{ij}-\nabla^{(3)}_i N_j - \nabla^{(3)}_j N_i \right).
\end{align}
Действие \eqref{GR_ADM_action} не содержит временных производных от функции хода и вектора сдвига. Как следствие, на уровне действия можно видеть, что из десяти независимых переменных четыре -- функция хода и вектор сдвига -- не являются динамическими и не могут быть признаны степенями свободы\footnote{Иными словами, вложение 3-мерной поверхности в 4-мерное пространство-время полностью определено собственной геометрией поверхности.}. Соответствующие уравнения Лагранжа генерируют четыре уравнения связи, которые исключат ещё четыре степени свободы. Этот вывод напрямую подтверждается гамильтонианом ОТО, который сводится к сумме Лагранжевых множителей:
\begin{align}
  H =\dot{\gamma}_{mn} \pi^{mn} - L = N\left[ 16\pi G~ G_{mnab} \pi^{mn} \pi^{ab} - \cfrac{\sqrt{\gamma} R^{(3)}}{16\pi G} \right] +N_m \nabla_n \pi^{mn}~.
\end{align}
Уравнения связи, генерируемые гамильтонианом, имеют следующий вид:
\begin{align}\label{constraint_operators}
  \begin{split}
    \mathcal{H}&=16\pi G ~G_{mnab} \pi^{mn} \pi^{ab} - \cfrac{ \sqrt{\gamma} R^{(3)} }{16\pi G}=0~, \\
    \mathcal{H}_n&=\nabla_m \pi^{mn} =0~.
  \end{split}
\end{align}

По полной аналогии с другими калибровочными системами гамильтониан ОТО тождественно равен нулю на уравнениях связи. В контексте теории гравитации это обстоятельство также указывает на проблему времени, так как полный гамильтониан системы должен быть связан с оператором сдвига по времени через уравнение Гейзенберга. Эту проблему можно разрешить формально определив оператор сдвига по времени через оператор $\hat{\mathcal{H}}$. Так производная по времени от любого оператора $\hat{\mathcal{O}}$ будет дана следующим аналогом уравнения Гейзенберга:
\begin{align}
  i \cfrac{\pd}{\pd t}~ \widehat{\mathcal{O}} = [\widehat{\mathcal{H}},\widehat{\mathcal{O}}]~.
\end{align}

Следующая проблема, возникающая на пути реализации программы канонического квантования, состоит в сложности определения пространства состояний. По полной аналогии с калибровочными теориями, мы должны ограничить пространство состояний, установив квантовые калибровочные условия. В случае гравитации эти условия генерируются классическими уравнениями связи \eqref{constraint_operators}:
\begin{align}
  \begin{split}
    \hat{\mathcal{H}}\ket{\Psi} &= 0 ~, \\
    \hat{\mathcal{H}}_n \ket{\Psi} &=0~.
  \end{split}
\end{align}
Первое уравнение системы называется уравнением Уилера-ДеВитта. В координатном представлении оно имеет следующий вид:
\begin{align}
  \left[-16 \pi G ~G_{mnab} \cfrac{\delta}{\delta \gamma^{mn}} \cfrac{\delta}{\delta \gamma^{ab}}-\cfrac{\sqrt{\gamma} R^{(3)}}{16\pi G}\right]\ket{\psi}=0~.
\end{align}
Это уравнение второго порядка по вариационным производным, что делает его аналогичным уравнению Клейна-Гордона-Фока, если в качестве аналога временной переменной рассматривать трёхмерную метрику. Можно было ожидать, что подобно уравнению Клейна-Гордона-Фока решения уравнения Уилера-ДеВитта можно разбить на два класса, аналогичные гармоникам с положительными и отрицательными частотами. Однако, такое разделение невозможно, что приводит к неустойчивости решений уравнения Уилера-ДеВитта. Связано это с тем, что аналог потенциального члена, содержащий трёхмерную кривизну, не ограничен снизу и может принимать сколь угодно отрицательные значения \cite{kuchar_general_1981,kuchar_conditional_1982} (подробное обсуждение приведено в \cite{kuchar_time_2011,RN19}). Следствием такого свойства решений уравнения Уилера-ДеВитта является невозможность создания пространства состояний с положительно определённой нормой.

Суммировать проблемы канонического подхода к квантовой гравитации следует следующим образом. Во-первых, аналогично другим калибровочным теориям каноническая квантовая теория гравитации изначально формулируется в переопределенном базисе. Собственные состояния оператора метрики не могут служить подходящим базисом, так как метрика не определена однозначно. Поэтому каноническая квантования теория нуждается в механизме, однозначно определяющем базисные состояния. Во-вторых, программа канонического квантования не может однозначно разрешить проблему времени. Классическая ОТО не имеет выделенной параметризации и, как следствие, от квантовой теории гравитации следует ожидать динамического возникновения времени. Наконец, квантовые уравнения связи не позволяют удовлетворительно определить пространство состояний. А именно, структура уравнения Уилера-ДеВитта не позволяет ввести положительно определённое скалярное произведение на множестве своих решений, что приводит к существованию в теории состояний с отрицательной нормой. В следствии этих фундаментальных недостатков программа канонического квантования гравитации в исходном виде не может быть признана полностью удовлетворительной.

\section{Эффективная квантовая гравитация}\label{effective_quantum_gravity}

Несмотря на все сложности, связанные с построением квантовой теории гравитации, существует формализм, позволяющий учесть эффекты, связанные с квантовой природой гравитационного поля. Это формализм эффективной теории поля, разработанный в рамках пертурбативного подхода к квантовой теории поля.

Существует два принципиальных подхода к эффективной теории поля, первый традиционно называется Вилсоновским в честь Кеннета Вилсона \cite{Wilson:1971bg,Wilson:1973jj}, чьи работы послужили основой для создания формализма эффективной теории поля \cite{Keller:1991cj,Kim:1998wz}; второй мы будем называть подходом Коулмана-Вайнберга по именам авторов, впервые изложивших его в статье \cite{Coleman:1973jx}. Детальное обсуждение обоих методов лежит за пределами этого обзора, так как техника эффективной теории поля распространена далеко за пределами теории гравитации \cite{Georgi:1994qn}. Применение метода эффективной теории поля в гравитации подробно обсуждено в \cite{Burgess:2003jk,Donoghue:2012zc,Donoghue:1994dn}.

Вернёмся к изучению системы полей $\{\Psi\}$ в рамках пертурбативного подхода к квантованию. Каждое поле содержит фоновую часть $\overline\Psi$ и малые возмущения фона $\psi$:
\begin{align}
  Z[~\overline\Psi ~] &= \int\mathcal{D}[\psi] \exp\left[ i \mathcal{A}\left[~\overline\Psi + \psi~\right]\right] \\
  &=\int\mathcal{D}[\psi]~\exp\left[i \mathcal{A}[~\overline\Psi] + i\cfrac{\delta\mathcal{A}[~\overline{\Psi} ~]}{\delta\overline\Psi}~ \psi + i\cfrac{\delta^2\mathcal{A}[~\overline\Psi~]}{\delta{\overline\Psi}^2}~ \psi^2 + O(\psi^3)  +\cdots\right] ~. \nonumber
\end{align}
В рамках пертурбативного подхода мы считаем, что фоновые поля $\overline\Psi$ первичны по отношению к малым возмущениям $\psi$ в том смысле, что именно фоновые поля определяют свойства малых возмущений.

Вилсоновский подход к эффективной теории поля состоит в построении описания данной системы полей в режиме малых энергий. Прежде всего, выделяется масштаб энергии $\Lambda$, определяющий границу между высоко- и низкоэнергетическими процессами. Все степени свободы в системе $\{\Psi\}$ разбиваются на лёгкие и тяжёлые по отношению к масштабу $\Lambda$, после чего все тяжёлые степени свободы исключаются из начальных и конечных состояний. Полученная теория описывает взаимодействие лишь лёгких степеней свободы, но тяжёлые степени свободы по-прежнему оказывают на неё влияние на уровне радиационных поправок. Таким образом из системы лёгких $\{\Psi_L\}$ и тяжёлых $\{\Psi_H\}$ степеней свободы, мы исключили все тяжёлые степени свободы и получили эффективное действие $\Gamma$, описывающее динамику лёгких степеней свободы при энергиях много меньших $\Lambda$:
\begin{align}
  \int\mathcal{D}[\Psi_L]~\mathcal{D}[\Psi_H]~\exp\left[i \mathcal{A}[\Psi_L,\Psi_H]\right] = \int \mathcal{D}[\Psi_L] ~\exp\left[i~\Gamma[\Psi_L]\right]~.
\end{align}
Этот подход получил широкое распространение в физике элементарных частиц, так как он позволяет согласовано учесть влияние тяжёлый степеней свободы стандартной модели, например, слабых калибровочных бозонов, на лёгкие степени свободы, например, на электроны \cite{Georgi:1990um,Fradkin:1984pq,Kim:1998wz}.

Подход Коулмена-Вайнберга, напротив, не требует исключения степеней свободы, а изучает макроскопическую динамику системы полей, генерируемую квантовыми взаимодействиями. В рамках пертурбативного подхода считается, что динамика малых возмущений системы определяется фоном системы, в то время как фон остаётся независимым от малых возмущений. Подход Коулмена-Вайнберга предлагает рассматривать малые квантовые возмущение как фундаментальный объект, порождающий классический фон. Иными словами, в генерирующем функционале следует произвести интегрирование по всем малым возмущениям, чтобы получить эффективное действие $\Gamma\left[\,\overline\Psi\,\right]$ описывающее динамику классического фона:
\begin{align}
  Z\left[ ~\overline\Psi ~ \right] = \int\mathcal{D}[\psi] \exp\left[i \mathcal{A}\left[\overline\Psi+\psi\right]\right] \overset{\text{def}}{=} \exp\left[i~ \Gamma[~\overline\Psi~]\right]~.
\end{align}

Эту логику можно продемонстрировать при помощи оригинальной работы \cite{Coleman:1973jx}, где рассматривалось скалярное поле с четырёхчастичным взаимодействием. На древесном уровне сектор взаимодействия модели описывает лишь четырёхчастичное взаимодействие. Однако, на уровне первой радиационной поправки сектор взаимодействия радикально меняется, так как радиационные поправки индуцируют новые многочастичные взаимодействия и, как следствие, динамически генерируют новый энергетический масштаб. Это находит своё отражение в том, что следующий ряд диаграмм может быть просуммирован, а его сумма привнесёт дополнительный вклад в сектор взаимодействия:
\begin{align}
  \begin{gathered}
    \begin{fmffile}{pic11}
      \begin{fmfgraph}(40,40)
        \fmfleft{i}
        \fmfright{o}
        \fmf{plain}{i,v,o}
        \fmffreeze
        \fmf{plain,right=1.5}{v,v}
      \end{fmfgraph}
    \end{fmffile}
  \end{gathered}
  +
  \begin{gathered}
    \begin{fmffile}{pic12}
      \begin{fmfgraph}(40,40)
        \fmfleft{i1,i2}
        \fmfright{o1,o2}
        \fmf{plain}{i1,a}
        \fmf{plain}{i2,b}
        \fmf{plain}{o1,x}
        \fmf{plain}{o2,y}
        \fmf{plain,tension=.7,left=.3}{a,b,y,x,a}
      \end{fmfgraph}
    \end{fmffile}
  \end{gathered}
  +
  \begin{gathered}
    \begin{fmffile}{pic13}
      \begin{fmfgraph}(40,40)
        \fmfleft{i1,i2,i3}
        \fmfright{o1,o2,o3}
        \fmf{plain}{i1,v1}
        \fmf{plain}{i2,v2}
        \fmf{plain}{i3,v3}
        \fmf{plain}{o1,v6}
        \fmf{plain}{o2,v5}
        \fmf{plain}{o3,v4}
        \fmf{plain,tension=1.31,left=.2}{v1,v2,v3,v4,v5,v6,v1}
      \end{fmfgraph}
    \end{fmffile}
  \end{gathered}
  +\cdots \to \cfrac12 ~\int\cfrac{d^4 k}{(2\pi)^4} ~\ln\left(1+ \cfrac{\lambda \phi^2}{2k^2}\right).
\end{align}
Таким образом, в полном согласии с указанной логикой, малые квантовые возмущения скалярного поля на уровне первой петлевой поправки индуцировали новое взаимодействие и изменили динамику классического поля.

Подход Коулмена-Вайнберга больше подходит для изучения гравитации, так как в рамках ОТО гравитоны являются безмассовыми и не могут задать выделенный масштаб энергии. Необходимо отметить, что в рамках других подходов возможно использование и Вилсоновского подхода. Например, в рамках теории струн (Виллоновское) эффективное действие для гравитации можно быть вычислено явно \cite{Alvarez-Gaume:2015rwa,Gross:1986mw,Fradkin:1985ys}. Однако, в наиболее общем случае, мы не должны ограничиваться одной теорией струн, а должны изучать наиболее общую форму эффективного действия.

Логика построения эффективных теорий поля позволяет восстановить вид эффективного действия исходя из соображений симметрии и размерности, даже если микроскопическое действие для системы неизвестно \cite{Burgess:2003jk}. В наиболее простом и практичном подходе к эффективной теории стоит утверждать, что эффективное действие для гравитации на уровне первой петлевой поправки состоит из всех операторов, генерируемых одноплетевыми диаграммами \cite{Barvinsky:1985an,Burgess:2003jk,Donoghue:2012zc,Donoghue:1994dn}, как следствие, имеет следующий вид:
\begin{align}\label{one_loop_effective_gravity_action}
  \Gamma_\text{1-loop}= \int d^4 x \sqrt{-g} \left[ -\cfrac{1}{16\pi G}~ R + c_1 R^2 + c_2 R_{\mu\nu}^2 \right].
\end{align}
Особо следует подчеркнуть, что расходящаяся часть эффективного действия как для гравитации, так и для других калибровочных теорий может быть восстановлена однозначно \cite{Barvinsky:1985an}.

Как мы отмечали в разделе \ref{quantum_GR_beyond_tree_level}, действие такого вида содержит духовые степени свободы. Однако, из-за того, что мы работаем в рамках эффективной теории, интерпретация этого результата должна быть принципиально иной. Тем более, что даже в рамках струнных моделей эффективное действие может содержать духовые степени свободы \cite{Alvarez-Gaume:2015rwa}. Только микроскопическое действие описывает фундаментальные степени свободы, а их динамика определяет вид эффективного действия. Как следствие, нет оснований считать, что новые полюса в пропагаторе возмущений, описываемых действием \eqref{one_loop_effective_gravity_action}, соотносятся с фундаментальными частицами. В частности, эффективное действие \eqref{one_loop_effective_gravity_action} генерируется, если использовать действие ОТО в качестве фундаментального. Полюс, соответствующий безмассовым возмущениям со спином $2$, должен быть связан с гравитоном, так как точно такой же полюс присутствует и в фундаментальном действии. Другие полюса, описывающие массивные возмущения спина $2$ и $0$, напротив, появляются только благодаря радиационным  поправкам, поэтому их нельзя интерпретировать как фундаментальные частицы.

Проще всего интерпретировать новые полюса как указание на то, что система входит в непертурбативный режим \cite{Solomon:2017nlh}. Неустойчивость и связанные с ней духовые состояния указывают лишь на то, что используемая нами теория возмущений теряет применимость. На это указывает также и то обстоятельство, что квазиклассические решения, описываемые эффективным действием вида \eqref{one_loop_effective_gravity_action} не имеют предела при $\hbar\to 0$ \cite{Simon:1990jn} (непертурбативны по $\hbar$).

Однако, новые полюса в пропагаторе, генерируемые эффективным действием \eqref{one_loop_effective_gravity_action} можно интерпретировать и как гравитационно связанные состояния \cite{Calmet:2014gya,Calmet:2015pea}. Эта интерпретация основывается на том, что духовые состояния, описываемые эффективной теорией, точно также как связанные состояния имеют отрицательную энергию. Вместе с этим, как показано в части \ref{quantum_GR_beyond_tree_level}, такие состояния возникают лишь на уровне первой радиационной поправки, то есть в режиме сильного поля, когда нелинейная природа гравитационного поля должна быть принята во внимание.

Важно подчеркнуть, что духовые состояния и связанная с ними неустойчивость может быть исключена из эффективной теории в нескольких особых случаях \cite{Solomon:2017nlh,Calmet:2018qwg}. В работе \cite{Calmet:2018qwg} было показано, что в эффективном действии для гравитации можно разделить степени свободы с разной массой и спином и привести эффективное действие к следующему виду (в калибровке Фейнмана):
\begin{align}\label{effective_gravity_1}
  \Gamma = \int d^4 x &\left[-\cfrac12 ~ \phi^{\mu\nu}\, \cfrac{C_{\mu\nu\alpha\beta}}{2} ~\square \phi^{\alpha\beta} +\cfrac12  ~ \psi^{\mu\nu} \cfrac{C_{\mu\nu\alpha\beta}}{2} ~\square \psi^{\alpha\beta} +\cfrac{m_2^2}{2} \, \left( \psi_{\mu\nu}^2- \psi^2\right) \right. \nonumber \\
    &\left. - \cfrac12~ \sigma (\square +m_0^2) \sigma +\kappa \left(\phi_{\mu\nu} + \psi_{\mu\nu} +\cfrac{1}{\sqrt{3}} ~ \sigma\right) T^{\mu\nu}  \right]~.
\end{align}
Тут $\phi_{\mu\nu}$, $\psi_{\mu\nu}$ и $\sigma$ соответственно безмассовые гравитоны, массивные духи со спином два и массивные скаляры; $T^{\mu\nu}$ обозначает тензор энергии-импульса материи, а массы степеней свободы определены формулой \eqref{gravity_eft_massess}.

Отметим, что мы исключили члены, описывающие взаимодействие гравитационных степеней свободы. Связано это с тем, что даже в рамках классической ОТО взаимодействие двух квадрупольных гравитационных волн может привести к образованию голой сингулярности \cite{khan_scattering_1971,pretorius_black_2018}. Следовательно, есть основания считать, что взаимодействие классических гравитационных степеней свободы является непертурбативным процессом и не может быть описано в рамках эффективной теории поля.

В работе \cite{Calmet:2018qwg} показано, что уравнения поля, генерируемые эффективным действием \eqref{effective_gravity_1}, совпадают с уравнениями поля, генерируемыми действием свободным от духовых состояний:
\begin{align}\label{effective_gravity_2}
  \Gamma = \int d^4 x &\left[-\cfrac12 ~ \phi^{\mu\nu} \cfrac{C_{\mu\nu\alpha\beta}}{2} ~\square \phi^{\alpha\beta} -\cfrac12  ~ \psi^{\mu\nu} \cfrac{C_{\mu\nu\alpha\beta}}{2} ~ \square\,\psi^{\alpha\beta} -\cfrac{m_2^2}{2} \, \left( \psi_{\mu\nu}^2- \psi^2\right) \right. \nonumber \\
    & \left. - \cfrac12~ \sigma (\square +m_0^2) \sigma +\kappa \left(\phi_{\mu\nu} - \psi_{\mu\nu} +\cfrac{1}{\sqrt{3}} ~ \sigma\right) T^{\mu\nu}  \right].
\end{align}
Несмотря на то, что это действие не содержит духовых состояний, оно описывает гравитационное отталкивание материи, так как член взаимодействия между массивными состояниями со спином $2$ имеет противоположный знак. Этот результат можно интерпретировать как указание на то, что в рамках эффективной теории поля описание состояний с отрицательной энергией неотличимо от описания состояний с положительной энергией, испытывающих гравитационное отталкивание.

Более того, есть основания утверждать, что духовые степени свободы не могут быть возбуждены в режиме низких энергий, где применим формализм эффективной теории поля. Так в работах \cite{Donoghue:1994dn,PhysRevD.67.084033,Iwasaki:1971vb} было показано, что потенциал взаимодействия двух тел с массами $M_1$ и $M_2$ в эффективной теории гравитации на уровне первой радиационной поправки дан следующим выражением:
\begin{align}
  V(r) = -\cfrac{G M_1 M_2}{r} \left[1+ 3 ~\cfrac{G(M_1+M_2)}{r} + \cfrac{41}{10\pi} \cfrac{G\hbar}{r^2} \right].
\end{align}
Поправки к стандартному потенциалу гравитационного взаимодействия подавлены гравитационной постоянной и спадают с расстоянием быстрее, чем $r$. Этот результат показывает, что радиационные поправки существенны для массивных тел и взаимодействий на малых расстояниях, то есть в режиме сильного поля, где описание гравитации методами эффективной теории поля может быть неприменимо.

Этот же вывод поддерживается результатами изучения гравитационного излучения двойных систем \cite{Calmet:2018qwg,Calmet:2018rkj,Calmet:2018uub}. В рамках эффективной теории поля мощность излучения двойной системы в нерелятивистском режиме получает дополнительный вклад от излучения массовых состояний со спином 2:
\begin{align}
  \left( \cfrac{dE}{d\omega} \right)_\text{massive} = \left( \cfrac{dE}{d\omega} \right)_\text{massless}  ~\theta(\omega-m_2)~ .  
\end{align}
Из-за того, что массивные и безмассовые степени свободы со спином $2$ имеют одну и ту же структуру взаимодействия с тензором энергии-импульса материи, они уносят из системы одинаковое количество энергии. Однако, из-за энергетической щели в спектре массивных мод их излучение начинается лишь при условии, что в системе достаточно энергии для их возбуждения. По этой причине в формуле для мощности излучения присутствует функция Хевисайда, определяющая критическую частоту системы. В нерелятивистском режиме частота вращения системы связана с расстоянием между компонентами $d$ следующим образом \cite{RN9,RN7}:
\begin{align}
  \omega^2= \cfrac{G(m_A + m_B)}{d^3} ~,
\end{align}
тут $m_A$ и $m_B$ -- массы компонентов системы. Используя консервативные эмпирические ограничения на $m_2$ \cite{Hoyle:2004cw} в работе \cite{Calmet:2018qwg} было показано, что система может начать излучать массивные моды лишь когда её компоненты сблизятся на расстояния порядка десятков сантиметров. Таким образом, массовые моды гравитационного поля будут возбуждаться лишь на последних этапах слияния двух чёрных дыр. Есть основания полагать, что в этом режиме взаимодействия центральных объектов двойных систем не могут быть описаны эффективной теорией поля, так как гравитация войдёт в режим сильной связи. Но даже если эффективная теория не потеряет свой применимости, есть основания полагать, что излучение массивных мод не окажет влияние на удалённого наблюдателя, так как сливающиеся объекты будут сокрыты горизонтом событий. 

Таким образом, часть сложностей, связанных с созданием квантовой теории гравитации, не возникает в рамках эффективной теории. Несмотря на то, что эффективная теория не может считаться фундаментальной и область её применимости ограничена режимом слабого поля, она позволяет получить верифицируемое описание гравитационных явлений, учитывающее квантовую природу взаимодействия (смотри уже перечисленные работы а также \cite{Latosh:2018xai,Arbuzov:2017nhg,Myers:2003fd,Goldberger:2004jt,Bjerrum-Bohr:2018xdl}). Кроме того, проблема духовых состояний, возникающая в квантовом описании гравитации, хоть и имеет место в рамках эффективной теории, но может быть существенно ослаблена. Есть основания полагать, что связанная с духовыми состояниями неустойчивость не реализуется в физически релевантных случаях. На это косвенно указывает и тот факт, что духи возникают и в эффективных теориях, генерируемых струнными моделями \cite{Alvarez-Gaume:2015rwa}. Поэтому эффективная квантовая гравитация является одним из наиболее простых и консервативных способов изучения квантовых гравитационных эффектов.

\section{Заключение}\label{conclusion}

В этом обзоре мы осветили базовые проблемы, возникающие при попытках построения квантовой теории гравитации. Мы ограничили обсуждение лишь двумя наиболее консервативными подходами и показали, что в их рамках построение квантовой теории гравитации невозможно при помощи стандартных методов.

Прежде всего мы обсудили пертурбативный подход к квантованию гравитации. Его суть состоит в создании квантовой теории малых возмущений гравитационного поля на классическом фоне. Как мы показали в разделе \ref{quantum_GR}, такой подход позволяет построить согласованную квантовую модель на древесном уровне. Проблемы пертурбативного подхода возникают при попытках расширить теорию на уровень первой радиационной поправки. Как мы обсудили в разделе \ref{quantum_GR_beyond_tree_level}, радиационные поправки к собственной энергии гравитона делают теорию неперенормируемой, а именно, для её перенормировки требуются контрчлены, отсутствующие в исходном лагранжиане. Вследствие чего стандартный метод перенормировки квантовой теории не может быть применён, а результаты вычислений не могут быть согласовано интерпретированы. При попытке построения перенормируемой квантовой теории неизбежно возникают духовые состояния. Духовые состояния появляются в спектре модели из-за членов с высшими производными, введёнными для улучшения свойств перенормируемости. Формальная причина возникновения этих членов связана со структурой гравитационного взаимодействия, которое квадратично по импульсам. Как мы показали в разделе \ref{quantum_GR_beyond_tree_level}, простой размерный анализ показывает необходимость введения высших производных в пертурбативную квантовую модель. Таким образом, расширить пертурбативное квантовое описание гравитации на уровень первых радиационных поправок невозможно при помощи стандартных методов перенормируемой квантовой теории поля.

В разделе \ref{nonperturbative_quantum_gravity} мы обратились к каноническому квантованию гравитации. Одним из важнейших недостатков программы квантования является проблема времени. При каноническом квантовании систем без гравитационного взаимодействия время является внешним по отношению к системе параметром эволюции. При каноническом квантовании гравитации принципиально невозможно установить время как внешний параметр, вследствие чего необходимо заново определить представление о времени как о динамически возникающей величине. На данный момент нет общепризнанного разрешения проблемы времени, поэтому мы использовали подход, называемый введение времени до квантования. Аналогично другим калибровочным теориям программа канонического квантования приводит к переопределенному базису состояний. Однако, если мы попробуем устранить эту проблему стандартным методом при помощи введения квантовых условий связи -- уравнений Уилера-ДеВитта -- то не получим удовлетворительный результат. Уравнение Уилера-ДеВитта аналогично уравнению Клейна-Гордона-Фока, так как оно является уравнением в вариационных производных второго порядка. Однако, в отличии от уравнения Клейна-Гордона-Фока, на решениях уравнения Уилера-ДеВитта невозможно ввести положительно определённую норму. Как следствие, пространство состояний неизбежно будет содержать состояния с отрицательной нормой, которые не могут быть интерпретированы в рамках стандартного подхода. Таким образом, программа канонической квантовой гравитации в исходной формулировке также не приводит к приемлемому результату.

Наконец, мы обсудили наиболее простой метод согласованного учёта квантовых эффектов в теории гравитации -- эффективную квантовую гравитацию, основанную на методе эффективного действия, разработанного в рамках физики частиц. Этот метод позволяет установить вид эффективного действия, определяющего динамику возмущений гравитационного поля в режиме малых энергий. Так как нам неизвестна истинная теория квантовой гравитации, вычислить эффективное действие невозможно, однако его вид можно восстановить исходя из наиболее общих соображений симметрии. Эффективное действие для квантовой гравитации на уровне первых радиационных поправок описывает духовые степени свободы, однако в рамках метода эффективной теории поля их существование является менее проблематичным. Прежде всего, духовые степени свободы, описываемые эффективным действием, не являются фундаментальными степенями свободы теории. Более того, как мы показали в разделе \ref{effective_quantum_gravity}, при определённых условиях эффективное действие для гравитации, содержащее духовые степени свободы, генерирует те же уравнения поля, что и действие свободное от духов, но содержащее гравитационное отталкивание. Иными словами, отрицательную энергию духовые степеней свободы можно интерпретировать как проявление гравитационного отталкивания. Также стоит отметить, что духовые степени свободы эффективной теории могут не быть возбуждены в низкоэнергетических процессах. В частности, прагматические оценки показывают, что духовые степени свободы в процессах слияния чёрных дыр возбуждаются лишь когда расстояние между компонентами системы достигает десятков сантиметров. Такие расстояния достигаются на последних этапах слияния, когда гравитационное поле может находиться в режиме сильной связи, что сделает эффективную теорию поля неприменимой, а сами гравитирующие объекты будет скрыты от удалённого наблюдателя горизонтом событий. Таким образом, в рамках эффективной теории поля духовые степени свободы могут быть контролируемы, а связанные с ними нестабильности могут не быть возбуждены в реалистичных условиях.

В разделах \ref{quantum_GR},\ref{quantum_GR_beyond_tree_level} и \ref{nonperturbative_quantum_gravity} мы описывали самые базовые трудности, возникающие при попытке построения квантовой теории гравитации. Современные методы исследования квантовой гравитации, такие как асимптотическая безопасность, AdS/CFT соответствие, супергравитация, теория струн и петлевая квантовая гравитация, предоставляют различные методы разрешения этих проблем. Однако, несмотря на прогресс, достигнутый в области квантовой теории гравитации, общепризнанная квантовая теория гравитационного взаимодействия ещё не создана. Поэтому все освещённые в этом обзоре проблемы ожидают своего разрешения.

\appendix
\appendixpage

\section{Формулы для малых метрических возмущений}\label{appendix_metric_perturbations}

Для метрики пространства-времени, данной формулой \eqref{metric_with_perturbations}, справедлив следующий набор формул. Если возмущения метрики $h_{\mu\nu}$ малы по сравнению с $\kappa$ эти выражения можно считать точными, а соответствующие ряды сходящимися. Если возмущения $h_{\mu\nu}$ не малы, то данные формулы можно последовательно рассматривать лишь как формальные выражения. Заметим также, что некоторые формулы, содержащие разложения до высших порядков, приведены в \cite{Goroff:1985th}.

\begin{align}
  g^{\mu\nu} =& \bar{g}^{\mu\nu} + \sum\limits_{n=1}^\infty (-\kappa)^n (h^n)^{\mu\nu}~, \\
  \sqrt{-g} =&\sqrt{-\bar{g}} \exp\left[ \cfrac12\operatorname{tr}\ln(1+\kappa~h_{\mu\nu}) \right] = \nonumber \\
  =&\sqrt{-g}\left[1+\cfrac{\kappa}{2} h -\cfrac{\kappa^2}{4} \left(h_{\mu\nu}^2 -\cfrac12 ~h^2\right) \right] + O(\kappa^3) ~, \nonumber \\
  \Gamma^\alpha_{\mu\nu} =&\bar{\Gamma}^\alpha_{\mu\nu} + \cfrac{\kappa}{2} \left(\nabla_\mu h_\nu{}^\alpha + \nabla_\nu h_\mu{}^\alpha - \nabla^\alpha h_{\mu\nu}\right)\nonumber \\
  &-\cfrac12 \sum\limits_{n=2}^\infty (-\kappa)^n (h^{n-1})^{\alpha\beta} \left(\nabla_\mu h_{\nu\beta} +\nabla_\nu h_{\mu\beta} - \nabla_\beta h_{\mu\nu}\right) ~, \nonumber \\
  \underline{\Gamma}^\alpha_{\mu\nu} =&\cfrac{\kappa}{2} \left[ \nabla_\mu h_\nu^\alpha + \nabla_\nu h_\mu^\alpha - \nabla^\alpha h_{\mu\nu} \right] ~, \nonumber \\
  \uunderline{\Gamma}^\alpha_{\mu\nu} =&-\cfrac{\kappa^2}{2} h^{\alpha\beta} \left[ \nabla_\mu h_{\nu\beta} + \nabla_\nu h_{\mu\beta} - \nabla_\beta h_{\mu\nu} \right] ~, \nonumber \\
  \underline{R}_{\mu\nu}{}^\alpha{}_\beta =& \nabla_\mu \underline{\Gamma}^\alpha_{\nu\beta} - \nabla_\nu \underline{\Gamma}^\alpha_{\mu\beta}~, \nonumber \\
  \uunderline{R}_{\mu\nu}{}^\alpha{}_\beta =& \nabla_\mu \underline{\underline{\Gamma}}^\alpha_{\nu\beta} -\nabla_\nu \underline{\underline{\Gamma}}^\alpha_{\mu\beta} + \underline{\Gamma}^\alpha_{\mu\sigma} \underline{\Gamma}^\sigma_{\nu\beta} - \underline{\Gamma}^\alpha_{\nu\sigma} \underline{\Gamma}^\sigma_{\mu\beta} ~, \nonumber
\end{align}
\begin{align}
  \underline{R}_{\mu\nu}{}^\alpha{}_\beta &= \cfrac{\kappa}{2} \left[[\nabla_\mu,\nabla_\mu] h^\alpha_{~\beta} - \nabla_\mu\nabla^\alpha h_{\nu\beta} - \nabla_\nu\nabla_\beta h_{\mu}^{~\alpha} + \nabla_\nu\nabla^\alpha h_{\mu\beta}+ \nabla_\mu\nabla_\beta h_\nu^{~\alpha} \right]~,  \nonumber \\
  \underline{R}_{\mu\nu} &\overset{\text{def}}{=} \underline{R}_{\sigma\mu}{}^\sigma_{~\nu} = \cfrac{\kappa}{2} \left[ \nabla^\sigma\nabla_\mu h_{\sigma\nu} + \nabla^\sigma\nabla_\nu h_{\sigma\mu} - \nabla_\mu\nabla_\nu h - \square h_{\mu\nu} \right]~ ,  \\
      \underline{R} &\overset{\text{def}}{=} \underline{R_{\mu\nu} g^{\mu\nu}} = \kappa[\nabla^{\mu\nu} h_{\mu\nu} - \square h] - \kappa \bar{R}_{\mu\nu} h^{\mu\nu} ~. \nonumber
\end{align}

\section{Операторы Риверса-фон Ньюенхайзена}\label{The_Operators}

Операторы Риверса-фон Ньюенхайзена определены в работах \cite{Rivers1964,VanNieuwenhuizen:1973fi}. Мы приведём выражения из работы \cite{Accioly:2000nm} более удобны для наших целей:
\begin{align}
  \begin{split}
    P^1_{\mu\nu\alpha\beta} &=\cfrac12 \left( \Theta_{\mu\alpha} \omega_{\nu\beta} + \Theta_{\mu\beta}\omega_{\nu\alpha} + \Theta_{\nu\beta}\omega_{\mu\alpha} + \Theta_{\nu\alpha} \omega_{\mu\beta}\right) ~, \\
    P^2_{\mu\nu\alpha\beta} &=\cfrac12 \left( \Theta_{\mu\alpha} \Theta_{\nu\beta} + \Theta_{\mu\beta} \Theta_{\nu\alpha}\right) -\cfrac13~ \Theta_{\mu\nu} \Theta_{\alpha\beta} ~,  \\
    P^0_{\mu\nu\alpha\beta} &=\cfrac13~ \Theta_{\mu\nu} \Theta_{\alpha\beta} ~, \\
    \bar{P}^0_{\mu\nu\alpha\beta} &= \omega_{\mu\nu} \omega_{\alpha\beta} ~, \\
    \bar{\bar{P}}^0_{\mu\nu\alpha\beta} &= \Theta_{\mu\nu} \omega_{\alpha\beta} + \Theta_{\alpha\beta} \omega_{\mu\nu}~.
  \end{split}
\end{align}
Тут мы использовали следующие обозначения:
\begin{align}
  \begin{split}
    \Theta_{\mu\nu}&=\eta_{\mu\nu} -\cfrac{k_\mu k_\nu}{k^2}~,\\
    \omega_{\mu\nu}&=\cfrac{k_\mu k_\nu}{k^2} ~.
  \end{split}
\end{align}
Операторы Риверса-фон Ньюенхайзена удовлетворяют следующей алгебре:
\begin{align}
  P^0_{\mu\nu\rho\sigma} P^1{}^{\rho\sigma\alpha\beta} &=0 ~, & P^0_{\mu\nu\rho\sigma} P^0{}^{\rho\sigma\alpha\beta} &= P^0_{\mu\nu}{}^{\alpha\beta} ~, \\
  P^1_{\mu\nu\rho\sigma} P^2{}^{\rho\sigma\alpha\beta} &=0 ~, & P^1_{\mu\nu\rho\sigma} P^1{}^{\rho\sigma\alpha\beta} &= P^1_{\mu\nu}{}^{\alpha\beta} ~,\nonumber \\
  P^0_{\mu\nu\rho\sigma} P^2{}^{\rho\sigma\alpha\beta} &=0 ~, & P^2_{\mu\nu\rho\sigma} P^2{}^{\rho\sigma\alpha\beta} &= P^2_{\mu\nu}{}^{\alpha\beta} ~, \nonumber
\end{align}
\begin{align}
  \bar{\bar{P}}^0_{\mu\nu\rho\sigma} P^0{}^{\rho\sigma\alpha\beta} &= \omega_{\mu\nu} \Theta^{\alpha\beta} ,& \bar{P}^0_{\mu\nu\rho\sigma} P^0{}^{\rho\sigma\alpha\beta} &=0 , & \bar{P}^0_{\mu\nu\rho\sigma} \bar{P}^0{}^{\rho\sigma\alpha\beta}&= \omega_{\mu\nu} \omega^{\alpha\beta} , \nonumber \\
  \bar{\bar{P}}^0_{\mu\nu\rho\sigma} P^1{}^{\rho\sigma\alpha\beta}  &=0 , & \bar{P}^0_{\mu\nu\rho\sigma} P^1{}^{\rho\sigma\alpha\beta} &=0 , & \bar{\bar{P}}^0_{\mu\nu\rho\sigma} \bar{\bar{P}}^0{}^{\rho\sigma\alpha\beta} &= \Theta_{\mu\nu} \Theta^{\alpha\beta} + 3 \omega_{\mu\nu} \omega^{\alpha\beta} \nonumber \\
  \bar{\bar{P}}^0_{\mu\nu\rho\sigma} P^2{}^{\rho\sigma\alpha\beta} &=0 , &\bar{P}^0_{\mu\nu\rho\sigma} P^2{}^{\rho\sigma\alpha\beta} &=0, & \bar{P}^0_{\mu\nu\rho\sigma} \bar{\bar{P}}^0{}^{\rho\sigma\alpha\beta} &= \omega_{\mu\nu} \Theta^{\alpha\beta} . \nonumber
\end{align}

\section{Пропагатор квадратичной гравитации}\label{quadratic_gravity_propagator}

Обратимся к квадратичной гравитации, данной следующим действием:
\begin{align}
  \mathcal{A}_\text{Stelle} =\int d^4 x \sqrt{-g} \left[ -\cfrac{2}{\kappa^2}\,R +c_1 \, R^2 + c_2 \, R_{\mu\nu}^2 \right] ~.
\end{align}
Покажем, что оно описывает духовые степени свободы над плоским про\-стран\-ством\--времинем.

Выделим из действие часть, квадратичную по возмущениям:
\begin{align}
  \mathcal{A}_\text{Stelle} =\int d^4 x \left[ -\cfrac12 ~ h^{\mu\nu} \left(\mathcal{O}_\text{Stelle}\right)_{\mu\nu\alpha\beta} \, \square h^{\alpha\beta} + O\left(h^3_{\mu\nu}\right) \right] ~.
\end{align}
Оператор $\mathcal{O}_\text{Stelle}$ в импульсном представлении выражается через операторы Риверса-фон Ньюенхайзена следующим образом:
\begin{align}
  \left(\mathcal{O}_\text{Stelle}\right)_{\mu\nu\alpha\beta} = \left(1+\cfrac{c_2\, \kappa^2}{2}\,k^2\right)\, P^2_{\mu\nu\alpha\beta} - 2 \left(1-(3c_1+c_2)\kappa^2 k^2\right)\, P^0_{\mu\nu\alpha\beta} ~.
\end{align}
Введём фиксирующий калибровку Фейнмана член:
\begin{align}
  \begin{split}
    \mathcal{A}_\text{gf} =& \int d^4 x \left(\pd_\mu h^{\mu\nu} - \cfrac12\,\pd^\nu h\right)^2 = \int d^4 x \left[-\cfrac12\, h^{\mu\nu} \left(\mathcal{O}_\text{gf}\right)_{\mu\nu\alpha\beta} \square h^{\alpha\beta} \right] ~, \\
    & \left(\mathcal{O}_\text{gf}\right)_{\mu\nu\alpha\beta}=P^1_{\mu\nu\alpha\beta}+\cfrac32\,P^0_{\mu\nu\alpha\beta} +\cfrac12\,\bar{P}^0_{\mu\nu\alpha\beta}-\cfrac12\, \bar{\bar{P}}^0_{\mu\nu\alpha\beta} ~.
  \end{split}
\end{align}

Таким образом, действие с фиксированной калибровкой имеет следующий вид:
\begin{align}
  \mathcal{A}_\text{Stelle} = \int d^4 x \sqrt{-g}\left[-\cfrac12\, h^{\mu\nu} \left(\mathcal{O}_\text{Stelle+gf}\right)_{\mu\nu\alpha\beta} \,\square h^{\alpha\beta} + O\left(h^3_{\mu\nu}\right)\right] ~,
\end{align}
\begin{align}
  & \left(\mathcal{O}_\text{Stelle+gf}\right)_{\mu\nu\alpha\beta}= \left(1+\cfrac{c_2\kappa^2}{2}\,k^2\right) P^2_{\mu\nu\alpha\beta}+\left(-\cfrac12+2 \kappa^2 k^2 \, (3c_1+c_2)\right)P^0_{\mu\nu\alpha\beta} \nonumber \\
  & +P^1_{\mu\nu\alpha\beta}+\cfrac12\,\bar{P}^0_{\mu\nu\alpha\beta}-\cfrac12\, \bar{\bar{P}}^0_{\mu\nu\alpha\beta} ~.
\end{align}
Оператор $\mathcal{O}_\text{Stelle+gf}$ обратим, с его помощью можно записать исходное действие в следующем виде:
\begin{align}
  &\mathcal{A}_\text{Stelle+gf} = \int\cfrac{d^4k}{(2\pi)^4} \left\{\cfrac12\, h^{\mu\nu} \mathcal{G}^{-1}_{\mu\nu\alpha\beta} h^{\alpha\beta}+O\left(h^3_{\mu\nu}\right) \right\} ~, \\
  &\mathcal{G}_{\mu\nu\alpha\beta} = \cfrac12\cfrac{C_{\mu\nu\alpha\beta}}{k^2}-\cfrac{P^2_{\mu\nu\alpha\beta}}{k^2-m_2^2}+\cfrac12\,\cfrac{P^0_{\mu\nu\alpha\beta}+3 \bar{P}^0_{\mu\nu\alpha\beta}+\bar{\bar{P}}^0_{\mu\nu\alpha\beta}}{k^2-m_0^2} ~. \label{stelle_gravity_propagator_appendix}
\end{align}
Выражение \eqref{stelle_gravity_propagator_appendix} определяет вид пропагатора возмущений, описываемых этой теорией. Второй член \eqref{stelle_gravity_propagator_appendix} имеет неправильный знак и описывает духовые возмущения.

\bibliography{Quantum_Gravity_EFT_Bibliography}
\bibliographystyle{gost2008}

\end{document}